\documentclass[10pt,aps,prb,twocolumn,amsmath,amssymb,floats,showpacs,floatfix,superscriptaddress]{revtex4-2}

\usepackage{graphicx}
\usepackage{bm}
\usepackage{amsmath}
\usepackage{amsfonts}
\usepackage{amsbsy}
\usepackage{physics}
\usepackage{verbatim}
\usepackage{color}
\usepackage{soul}
\usepackage{titlesec}
\usepackage{indentfirst}
\usepackage[normalem]{ulem}
\usepackage[dvipsnames]{xcolor}
\titlespacing*{\section}{0pt}{15pt}{15pt}

\usepackage{placeins} 

\usepackage{caption}
\usepackage{subcaption}
\usepackage[urlcolor=blue,colorlinks=true,citecolor=blue,linkcolor=blue]{hyperref}
\usepackage[nameinlink]{cleveref}
\crefmultiformat{figure}{~#2#1#3}{,\!~#2#1#3}{, #2#1#3}{ and~#2#1#3}
\crefname{equation}{Eq.}{Eqs.}  
\crefname{figure}{Fig.}{Figs.} 
\crefname{section}{Sec.}{Secs.} 
\crefname{appendix}{Appendix}{Appendices}

\newcommand{\thore}[1]{\textcolor{blue}{#1}}

\DeclareMathOperator{\sgn}{sign}

\begin{document}

\title{Time-reversal symmetric topological superconductivity in Machida-Shibata lattices}
\author{Ioannis Ioannidis}
\affiliation{I. Institute for Theoretical Physics, University of Hamburg, D-22607 Hamburg, Germany}
\affiliation{Centre for Ultrafast Imaging, Luruper Chaussee 149, D-22761 Hamburg, Germany}
\author{Ching-Kai Chiu}
\affiliation{RIKEN Interdisciplinary Theoretical and Mathematical
Sciences (iTHEMS), Wako, Saitama 351-0198, Japan}
\author{Thore Posske}
\affiliation{I. Institute for Theoretical Physics, University of Hamburg, D-22607 Hamburg, Germany}
\affiliation{Centre for Ultrafast Imaging, Luruper Chaussee 149, D-22761 Hamburg, Germany}

\date{\today}
\begin{abstract}
 Recent experiments engineered special spin-degenerate Andreev bound states in atomic cages of adatoms on superconductors, the Machida-Shibata states, a promising building block for quantum matter. Here, we investigate the formation of time-reversal symmetric bands by hybridizing multiple such states and analyzing their electronic topological properties assuming small on-site electron repulsion. The low-energy theory shows that competing emerging singlet and triplet superconducting pairings drive the formation of topologically non-trivial phases in symmetry class DIII. We, therefore, predict Kramers pairs of Majorana zero modes appear at the ends of Machida-Shibata chains, while two-dimensional lattices host helical Majorana edge modes. Additionally, we discover extended regions in the Brillouin zone with vanishing superconducting pairings, which are lifted by the repulsive electron interactions. Our findings offer new perspectives for manipulating topological superconductivity and pairings in non-magnetic adatom systems.
\end{abstract}
\maketitle
\section{INTRODUCTION}
Topological superconductivity has been a subject of immense research motivated by the potential technological impact of topological protection and non-abelian effects~\cite{RevModPhys.80.1083,Leijnse_2012}, particularly in platforms that potentially host Majorana modes~\cite{Kitaev2001,Sarma2015,PhysRevLett.105.077001,PhysRevLett.105.177002,PhysRevLett.100.096407}. The initial momentum of the field was generated by the observation of zero-bias peaks in nanowires proximitized by superconductors~\cite{doi:10.1126/science.1222360,Das2012}, a path that recently faced challenges due to impurities leading to ambiguities~\cite{Zhang2021,PhysRevB.105.174519}. On the other hand, adatoms deposited on clean superconducting surfaces provide unprecedented control over system purity and hints for a plethora of time-reversal-symmetry breaking phases, including quantum spin systems~\cite{Liebhaber2022,PhysRevB.108.L220506}, topological nodal point superconductivity~\cite{Bazarnik2023}, localized Majorana zero modes in 1D artificial magnetic chains~\cite{doi:10.1126/science.1259327,doi:10.1126/sciadv.aar5251}, and propagating chiral Majorana modes in 2D magnet-superconductor hybrids~\cite{Ménard2017,doi:10.1126/sciadv.aav6600}. 

A central objective of this study is to propose a platform for realizing topological superconductivity in symmetry class DIII~\cite{PhysRevB.78.195125,10.1063/1.3149495,RevModPhys.88.035005}. Atoms with magnetic anisotropy, when deposited on superconductors, induce spin-polarized electronic Yu-Shiba-Rusinov (YSR) states~\cite{Yu1965Bound,Shiba1968classical,Rusinov1969theory} which serve as building blocks for constructing topological superconductors.  Yet, the realization of non-trivial topological phases in time-reversal-symmetric adatom systems remains experimentally elusive. For adatom species with negligible magnetic polarization the predicted electronic spin-degenerate in-gap states~\cite{1972PThPh..47.1817M} are typically energetically close to the coherence peaks and, thus, can typically be neglected~\cite{RevModPhys.78.373}.

However, recent experiments have emphasized the significance of such bound states, the so-called Machida-Shibata states (MSSs), by assembling adatoms in spatially tunable atomic cages and controlling their energy as particle-in-a-box states~\cite{Schneider2023,PhysRevB.110.L100505,Ton2025}, first demonstrated in Ag quantum corrals deposited on thin Ag(111) islands grown on superconducting Nb(110)~\cite{Schneider2023}. In principle, MSSs can be realized in all superconductor-metal or -alloy composites with a Shockley state extending over their surfaces~\cite{PhysRev.56.317,PhysRevB.75.195414,PhysRevLett.98.186807}. Additionally, manifestations of MSSs are reported in transmon qubit Josephson junctions~\cite{PhysRevLett.124.246802,PhysRevLett.124.246803}, and similar hybridized states have been realized in Andreev molecule setups~\cite{Su2017,Janvier2015}. The MSS, albeit spatially extended, constitutes a tunable non-magnetic analogue to the YSR state and opens possibilities for engineering distinct time-reversal-symmetric topological phases. 
\begin{figure}[t]
    \centering
    \includegraphics[width=1.\linewidth]{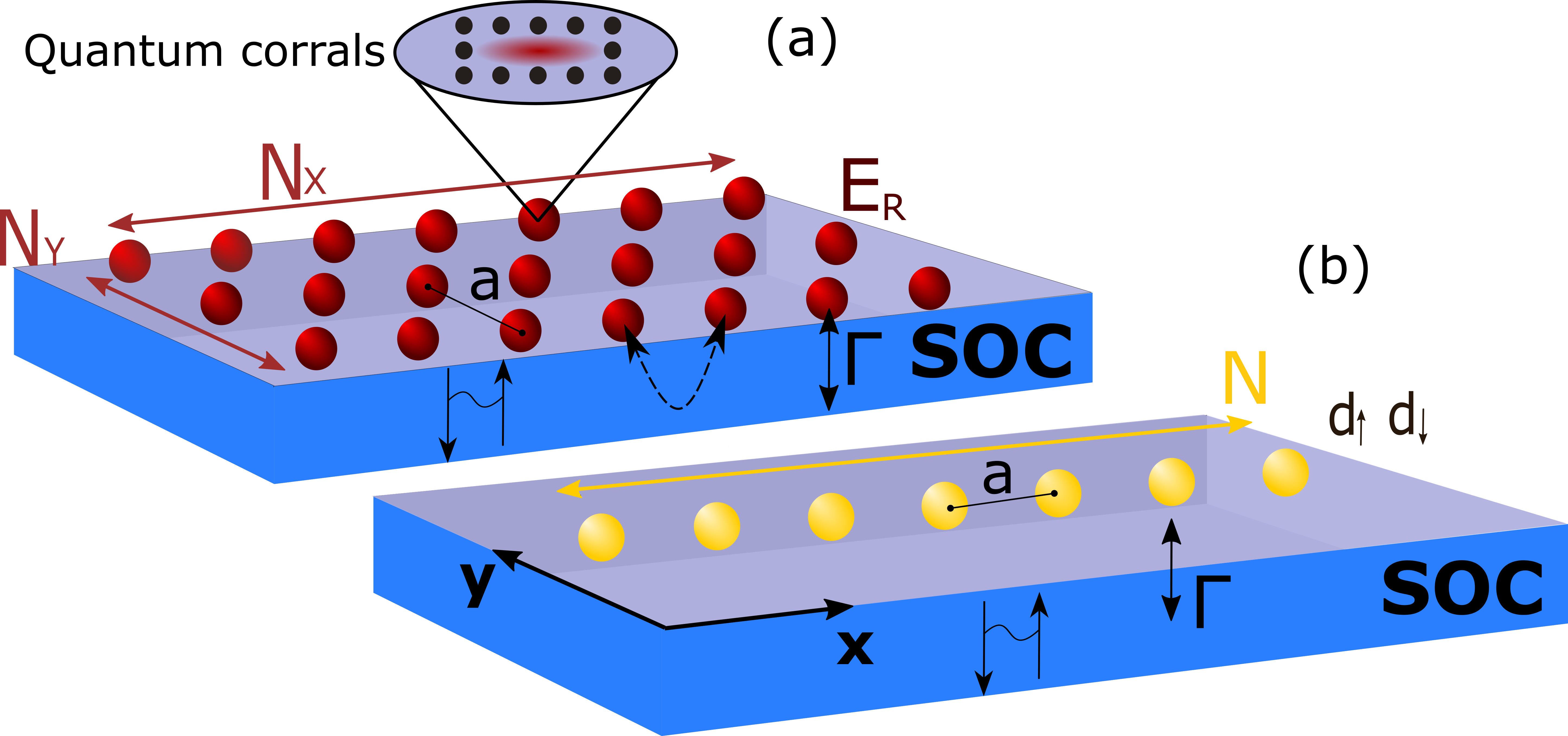}
    \caption{Setup: Adatom manipulation on the surface of a superconductor (blue) with Rashba spin-orbit coupling (SOC) constructs arrays of quantum corrals and induces Machida-Shibata states. Each state is described by localized spinful electronic levels (here called d-levels) with energy $E_{\rm R}$ (red and yellow spheres) which are coupled via a common bulk. Geometries: (a) Square lattice with lattice constant $a$ ($N_x\times N_y$ sites) and (b) One-dimensional $N-$site chain.}
    \label{Fig.:Schematics}
\end{figure}

In this manuscript, we investigate the topological electronic phases of
MS bands formed by coupling multiple MSSs via indirect tunneling
processes to a common bulk. To this end, we adapt
the Green’s function techniques developed for magnetic adatom systems~\cite{PhysRevLett.114.106801,PhysRevB.104.245133,Li2016}. Since
time-reversal symmetry is naturally preserved, the MS bands belong to
symmetry class DIII, in contrast to Shiba bands which belong to symmetry
class D due to the magnetic impurities~\cite{PhysRevB.88.155420}. Achieving topologically
nontrivial phases in class DIII by proximity-induced superconductivity
from the s-wave parent superconductor alone is impossible to achieve,  due to the no-go theorem of Ref.~\cite{PhysRevB.94.161110}. Yet, in the presence of small repulsive electron
interactions inside the corrals that break this prerequisite, we
demonstrate that the MS bands can intersect the Fermi energy, leading to
topologically nontrivial phenomenology. Regarding experimental signatures, the MS bands show no sensitivity to the spin-polarization of differential conductance STM measurements, in contrast to Shiba bands and accommodate distinct Kramers' protected double Majorana features in 1D and helical propagating Majorana modes in 2D. Non-trivial phases in this class require unconventional superconducting correlations and a SU(2)-spin symmetry breaking mechanism~\cite{PhysRevB.88.134523,PhysRevB.94.161110,PhysRevB.96.161407,PhysRevLett.121.196801}, which in our case is achieved by non-vanishing Rashba spin-orbit coupling (SOC) in the superconducting bulk. Physically, the Rashba SOC is motivated by the crystal inversion symmetry breaking in the direction perpendicular to the host material's surface in combination with spin-orbit coupling in the atomic levels~\cite{Bihlmayer2022}. For more complex materials, such as Bi/Ag(111) surface alloys, structural anisotropies on the confinement surface lead to additional unconventional Rashba bands and an emerging singlet-triplet superconducting order parameter competition~\cite{PhysRevB.79.245428,PhysRevB.76.073310,PhysRevB.110.134517}. 

We describe the microscopic theory and derive the effective low-energy four-band Hamiltonian to demonstrate a competition between the emergent singlet and triplet superconducting couplings which dictates the spectral gap closings and, thus, the topological phases of the system. Subsequently, we consider one- and two-dimensional lattices of MSSs, see \cref{Fig.:Schematics}, and picture the boundary modes and their spatial profiles. By tuning physical parameters such as the on-site MS energy and SOC strength, we construct phase diagrams that accommodate trivial and non-trivial phases. We investigate the effect of repulsive Coulomb interactions using a mean-field theory, and demonstrate that they are the essential element for making the induced singlet parameter non semi-definite and allowing non trivial phases to appear when all couplings are considered. Additionally, we find extended regions in the Brillouin zone where both the singlet and triplet superconducting order parameters become flat and vanish in 1D chains of MSSs when a 2D superconductor plays the role of the substrate. Notably, we observe cusp-like features in the effective pairing amplitudes, akin to those predicted in YSR systems~\cite{PhysRevB.88.155420}. Thereby, our work draws attention to the topological properties and applications of non-magnetic adatom lattices on superconductors.

\section{MICROSCOPIC DESCRIPTION}\label{Sec.Model}
 We describe the MSSs by a superconducting Anderson model for localized electrons. This neglects the spatial structure of the MSS inside the corrals but faithfully captures their essential characteristics, namely their energetic position and particle-hole asymmetry~\cite{Schneider2023,Ton2025}. We extend this approximation to multiple coupled quantum corrals on a lattice described by electron annihilation operators $\hat{d}_{j,\sigma}$ localized at positions $\textbf{R}_{j}$, which we refer to as d-levels, in the presence of superconductivity, see \cref{Fig.:Schematics}. To this end, we consider the following Hamiltonian in second quantization
\begin{equation}\label{eq:Hamiltonian}
\begin{split}
    &\hat{H}=\hat{H}_{\rm MS}+\hat{H}_{\rm SC}+ \hat{H}_{\rm T},\\
    &{\rm with} \ \hat{H}_{\rm MS}=E_{\rm R} \sum_{\sigma,j}d^{\dagger}_{\sigma,j} d_{\sigma,j}+U \sum_j d^{\dagger}_{\uparrow,j} d_{\uparrow,j}d^{\dagger}_{\downarrow,j} d_{\downarrow,j},\\
    &\hat{H}_{\rm SC}=\sum_{ \textbf{k},\sigma} \epsilon_{\textbf{k}} c^{\dagger}_{\textbf{k},\sigma} c_{\textbf{k},\sigma}+ \lambda \sum_{\textbf{k}}|\textbf{k}|\left(ie^{-i\theta(\textbf{k})}c^{\dagger}_{\textbf{k},\uparrow}c_{\textbf{k},\downarrow}+h.c\right)\\&-\Delta \sum_{\textbf{k}}\left( c^{\dagger}_{\textbf{k} \uparrow} c^{\dagger }_{-\textbf{k} \downarrow}+h.c\right)
    ,\\
    &\hat{H}_{\rm T}=V\sum_{\textbf{k},\sigma,j}\left(e^{i\textbf{k}\textbf{R}_j}c^{\dagger}_{\textbf{k},\sigma} d_{\sigma,j} + h.c\right),
\end{split}
\end{equation}
    where $\hat{H}_{\rm MS},\hat{H}_{\rm SC},\hat{H}_{\rm T}$ describe the corral levels, substrate, and the tunneling between them, respectively. Here, $\epsilon_{\textbf{k}}=\textbf{k}^2/2m-E_F$ is the dispersion relation of the continuum bulk superconductor with Fermi energy $E_F$, effective electron mass $m$, and SOC strength $\lambda$. We set $\hbar=1$. Throughout, $a$ denotes the lattice constant of the chain or square lattice of d-levels, as illustrated in \cref{Fig.:Schematics}. The dimensionality of the superconducting bulk in \cref{eq:Hamiltonian} is not constrained. To obtain analytical results for finite SOC, we focus on a 2D bulk superconductor~\cite{Heimes_2015}. $E_{\rm R}$ is the spin-independent energy of the d-levels while $\theta(\textbf{k})$ refers to the azimuthal angle of the wavevector $\textbf{k}$ in the plane parallel to the surface. Additionally, we consider local scattering at the MSS sites with potential $V_j(\textbf{r})=V \delta(\textbf{r}-\textbf{R}_j)$, where $\textbf{r}$ refers to the position of the bulk electrons. The on-site repulsive interaction, $U$, is qualitatively included to account for Coulomb repulsions that originate from the real-space confinement of the d-levels. This interaction is neglected first, i.e., $U=0$, until specified. In the case of a single pair of d-levels in \cref{eq:Hamiltonian}, a pair of MSSs appears in the superconducting gap~\cite{1972PThPh..47.1817M}. The broadening of the d-levels is determined by the hybridization strength $\Gamma=\pi \nu \abs{V}^2$, where $\nu$ is the normal state density of states of the bulk at the Fermi energy. A similar model has been studied in the context of dense magnetic adatom chains~\cite{PhysRevLett.114.106801}. In contrast, the model in \cref{eq:Hamiltonian} preserves time-reversal-symmetry and both spin-species need to be considered. 
\begin{figure*}[htbp]
    \centering
    \begin{subfigure}{0.365\textwidth}
    \centering
    \caption[]{}
    \includegraphics[width=\textwidth]{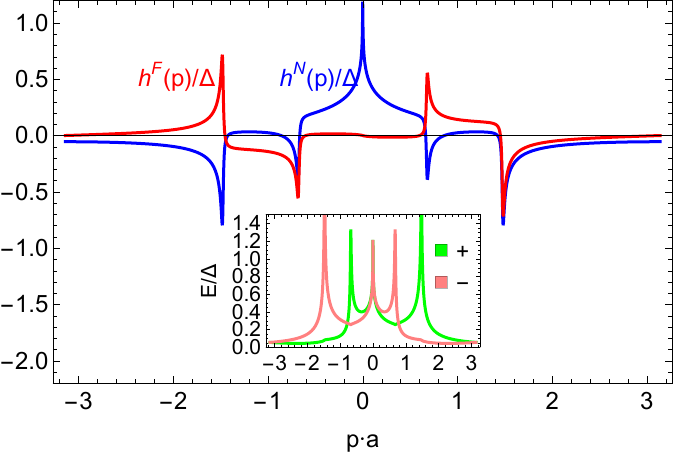}
      \label{Fig.:Chains-Effective Parameters a}
    \end{subfigure}
    \hfill
    \begin{subfigure}{0.365\textwidth}
    \centering
    \caption[]{}

\includegraphics[width=\textwidth]{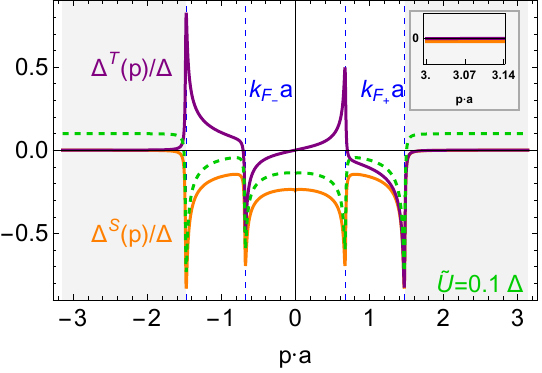}
    \label{Fig.:Chains-Effective Parameters b}
    \end{subfigure}
    \hfill
     \begin{subfigure}{0.25\textwidth}
    \centering
    \caption[]{}

\includegraphics[width=\textwidth]{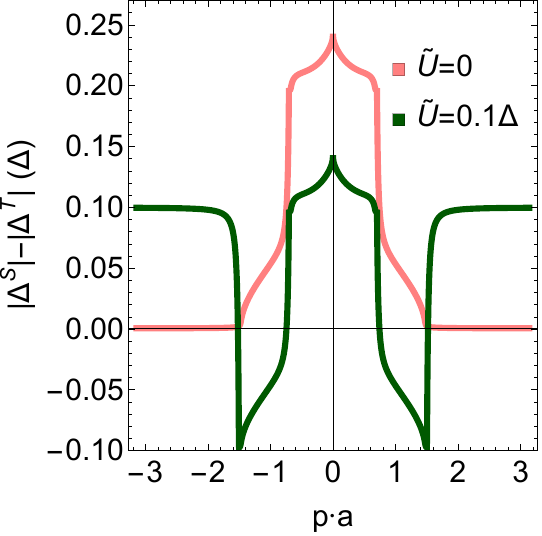}
    \label{Fig.:Chains-Effective Parameters c}
    \end{subfigure}
    \caption{Flattening of effective couplings in the low-energy theory for one-dimensional chains. (a) Long-range hoppings, $h^N(p)$, and spin-flip parameters, $h^F(p)$, in the first Brillouin zone for $E_{\rm R}=0$. The inset shows the two helicity bands of the dispersion relation. (b) Singlet, $\Delta^S(p)$ (orange for $\tilde{U}=0$ and dashed green for $\tilde{U}=0.1\Delta$), and triplet, $\Delta^T(p)$, superconducting order parameters. The inset zooms in the flat region (gray shading). (c) Difference $|\Delta^S|-|\Delta^T|$ of the singlet and triplet parameters in the presence ($\tilde{U}=0.1\Delta$) and absence ($\tilde{U}=0$) of interactions. (adopted parameter values: $k_{\rm F}a=1.2$, $\tilde{\lambda}=\lambda/u_{\rm F}=0.4$, $m=10^{-2} (a^2 \Delta)^{-1}$, $\abs{V}=5\Delta$ and $N_{\mathrm{cutoff}}=1000$.)}
    \label{Fig.:Chains-Effective Parameters}
\end{figure*}
 To study the topological phases of the system, it suffices to focus on the low-energy limit $E/\Delta\rightarrow0$~\cite{Wang_2013}, because topological phases can only change upon closure of the spectral gap. In this limit, we integrate out the bulk modes and derive an effective BdG Hamiltonian (see \cref{Appendix:Derivation} for details)
\begin{equation}\label{eq:bdgH}
\begin{split}
  H_{\rm eff}=&\begin{pmatrix} h_{i,j} & \Delta_{i,j}\\
 \Delta^{\dagger}_{i,j} &-\sigma_z h_{i,j} \sigma_z
    \end{pmatrix},\\
     h_{i,j}=&\begin{pmatrix}
          h^N_{i,j} & h^F_{i,j} \\ h^{F\dagger}_{i,j}
         & h^N_{i,j}
    \end{pmatrix},\Delta_{i,j}=\begin{pmatrix}
          \Delta^S_{i,j} & \Delta^T_{i,j} \\  \Delta^{T*}_{i,j}
         & -\Delta^S_{i,j}
    \end{pmatrix}
\end{split}
\end{equation}
defined in the basis $\begin{pmatrix}
         d^{\dagger}_{\uparrow,j} &
        d^{\dagger}_{\downarrow,j} &
        d_{\downarrow,j} &
        d_{\uparrow,j} 
    \end{pmatrix}^{\dagger},$ which describes a system of coupled MSSs and the corresponding bands. The matrix elements $h^N_{i,j}$ include the on-site energy of the d-levels and long-range hoppings mediated by the bulk. The $h^F_{i,j}$ matrix elements describe long-range spin-flips between  d-levels. The $\Delta^S_{i,j}$ and $\Delta^T_{i,j}$ matrix elements are the induced singlet (on-site and long-range) and triplet superconducting pairings, respectively. The induced matrix elements are proportional to the scattering strength $\Gamma$. Importantly, $h^F_{i,j}$ and $\Delta^T_{i,j}$ vanish for zero SOC $\lambda=0$, see \cref{Appendix:Derivation}.

For periodic boundary conditions, we transform the Hamiltonian in \cref{eq:bdgH} to momentum space
\begin{equation} \label{eq:momentumH}
     H_{\rm eff}(\textbf{p})= \begin{pmatrix}
            h^N(\textbf{p})&  h^F(\textbf{p})&  \Delta^S(\textbf{p})&\Delta^{T}(\textbf{p}) \\ 
         h^{F*}(\textbf{p}) &  h^N(\textbf{p}) &-\Delta^{T*}(\textbf{p}) & -\Delta^S(\textbf{p})\\
       \Delta^{S*}(\textbf{p})&-\Delta^{T}(\textbf{p})& - h^N(\textbf{p}) &h^F(\textbf{p})\\
      \Delta^{T*}(\textbf{p}) &- \Delta^{S*}(\textbf{p}) &  h^{F*}(\textbf{p})& - h^N(\textbf{p})
    \end{pmatrix},
\end{equation}
defined in the basis $\Psi_\textbf{p}=\begin{pmatrix}
         d_{\uparrow,\textbf{p}}^{\dagger}&
        d_{\downarrow,\textbf{p}}^{\dagger}&
        d_{\downarrow,-\textbf{p}}&
        d_{\uparrow,-\textbf{p}} 
    \end{pmatrix}^{\dagger} $ and $
  h^N(\textbf{p})= \sum_{\textbf{x}}  e^{i \textbf{p} \textbf{x}} h^N_\textbf{x}$, where $\textbf{x}$ runs over the entire lattice and similarly for $  h^F(\textbf{p}), \Delta^S(\textbf{p}), \Delta^T(\textbf{p})$. Importantly, $h^N(\textbf{p})$ and $\Delta^S(\textbf{p})$ are even but $ h^F(\textbf{p})$ and $\Delta^T(\textbf{p})$ are odd functions in momentum $\textbf{p}$. In the following, we systematically classify the topological phases associated with \cref{eq:momentumH} by studying its symmetry class and topological invariant~\cite{PhysRevB.78.195125,10.1063/1.3149495,RevModPhys.88.035005}. The effective Hamiltonian in \cref{eq:momentumH} is generally gapped due to the induced on-site singlet pairing, and a nonvanishing SOC breaks the spin-SU(2) symmetry. Inheriting from the original Hamiltonian in \cref{eq:Hamiltonian}, $H_{\rm eff}$ belongs to symmetry class DIII, respecting the particle-hole symmetry  $\mathcal{C}=\tau_x \sigma_x K$ with $\mathcal{C}^2=1$ and time-reversal symmetry $\mathcal{T}= i\tau_z\sigma_y K$ with $\mathcal{T}^2=-1$, where $\sigma,\tau$ are Pauli matrices in spin and particle-hole space, respectively. We conclude that the eigenmodes of \cref{eq:momentumH} appear in Kramers pairs. Without SOC, the system reduces to two spinless sub-systems in class BDI. For strictly translational invariant one-dimensional systems, the long-range spin-flip parameter $h^{F}(p)$ is purely real and the triplet pairing $\Delta^{T}(p)$ is purely imaginary (see Appendix~B). As a consequence, the unitary operator $\mathcal{U}=\tau_{z}\sigma_{y}$ commutes with $H_{\rm eff}(p)$. Combining $\mathcal{U}$ with the time-reversal and particle-hole operators yields two antiunitary symmetries $T'=T\mathcal{U}, C'=C\mathcal{U}$, which satisfy $(T')^{2} = +1$ and $(C')^{2} = -1$, placing the system in symmetry class~CI which carries a $\mathcal{Z}$ topological invariant. For the sake of generality, we keep our analysis at the level of the single DIII system. To calculate the single-particle eigenvalues, we write \cref{eq:momentumH} as a tensor product of Pauli matrices and square it twice, see \cref{Appendix:eigenvaluecalculation} for the derivation. Interestingly, for 1D lattices, the condition for a zero energy crossing, i.e., $E_{\pm}=0$, is equivalent to finding a momentum $p_0$, such that 
\setlength{\abovedisplayskip}{20pt}
\setlength{\belowdisplayskip}{20pt}
\begin{equation}
\begin{split}\label{eq:Condition1D}
\Delta^S(p_0)=&\pm \Im{\Delta^T(p_0)}, \\ h^N(p_0)=&\mp \Im{h^F(p_0)},
\end{split}
\end{equation} 
see \cref{Appendix:eigenvaluecalculation} for details. Once the first condition in \cref{eq:Condition1D} is satisfied, the second one can be fulfilled by an appropriate choice of the on-site energy $E_{\rm R}$, which can be freely tuned by the size of the corrals. Qualitatively, the competition between the singlet, $\Delta^S$, and triplet, $\Delta^T$, pairings determines the zero-energy crossings and, thus, the topological phases~\cite{PhysRevB.88.134523}.

\section{RESULTS}\label{Sec.Results}
Next, we demonstrate concrete examples by considering a 2D bulk superconductor with a linearized dispersion $\epsilon_{\textbf{k}_\pm}=\tilde{\upsilon}_F (k_{\pm}-k_{F_\pm})$ around the Fermi level, where $\pm$ corresponds to the two helicity bands in the presence of SOC. The bands are characterized by the Fermi wavevectors $k_{F_\pm}$, SOC renormalized Fermi velocity, $\tilde{\upsilon}_F$, and the modified normal state density-of-states $\nu_{\pm}$ at the Fermi energy, see \cref{Appendix:2DbulkSystem}. In this case, the matrix elements are 
\begin{equation}\label{eq:MatrixElements2D}
\begin{split}
 &h^{N}_{n,m}=E_{\rm R}\delta_{n,m}+(1-\delta_{n,m}){\rm \Im}\left(w^e_{n,m}\right),
    \\
    &h^{F}_{n,m}=(1-\delta_{n,m})e^{-i\phi_{n,m}}{\rm Re}\left(w^o_{n,m}\right),\\
    &\Delta^{S}_{n,m}=-\Gamma \delta_{n,m}-(1-\delta_{n,m}){\rm \Re}\left(w^e_{n,m}\right),\\
     &\Delta^{T}_{n,m}=-(1-\delta_{n,m})e^{-i\phi_{n,m}}{\rm Im}\left(w^o_{n,m}\right),\\
   &{\rm with} \ 
w^{e}_{n,m}=\sum_{\mu=\pm}\dfrac{\Gamma_{\mu}}{2}\left(J_{0}\left[x^{\mu}_{n,m}\right]+iH_{0}\left[x^{\mu}_{n,m}\right]\right),\\
&w^{o}_{n,m}=\sum_{\mu=\pm}\dfrac{\mu\Gamma_{\mu}}{2}\left(iJ_{1}\left[x^{\mu}_{n,m}\right]+H_{-1}\left[x^{\mu}_{n,m}\right]\right),
\end{split}
\end{equation}
where $J_n,H_n$ are the $n^{\rm th}$ Bessel and Struve functions, respectively, which are holomorphic in the whole complex plane for integer $n$. Also, $\phi_{n,m}$ is the azimuthal angle of the vector $\textbf{R}_n-\textbf{R}_m$ and $ x^\pm_{n,m}=\left(k_{F_{\pm}}+i\xi^{-1}\right)\abs{\textbf{R}_{n}-\textbf{R}_{m}}$, where $\xi$ is the superconducting coherence length. Here, the renormalized scattering strength, $\Gamma_{\pm}=\pi\nu_{\pm}\abs{V}^2$, is proportional to $\Gamma$. Importantly, all matrix elements in \cref{eq:MatrixElements2D}, except for the on-site energy $E_{\rm R}$, scale with the scattering strength $\Gamma$. Any modification in $\Gamma$ can be compensated by an appropriate adjustment to $E_{\rm R}$, guaranteeing that the value of $\Gamma$ (or V) is not critical for satisfying the condition $E_{\pm}=0$ and, thus, realizing non-trivial phases. In the case of a 1D lattice, this statement can be inferred from the matrix elements in \cref{eq:MatrixElements2D} since both $\Delta^S$ and $\Delta^T$ are proportional to $\Gamma$ and, thus, the first condition in \cref{eq:Condition1D} does not depend on $\Gamma$. Similarly, the effective mass $m$ rescales the renormalized SOC strength $\tilde{\lambda}=m\lambda/k_{\rm F}$ and $\nu_{\pm}$ and, thus, $\Gamma$ as well, see \cref{eq:supprenormalized} in the Appendix. Therefore, the only critical parameters for realizing non-trivial phases are the Fermi wavevector, the SOC strength, and the on-site energy. In the following, we fix the lattice positions, \{$\textbf{R}_j$\}, of the d-levels and study the cases of one-dimensional chains and two-dimensional square lattices separately while varying the aforementioned critical parameters. For the plots that follow, we fix the effective mass to be of the order of the electron mass, $m\sim m_e$, and consider the scattering strength $\abs{V}=5\Delta$, which leads to $\Gamma \sim 0.1\Delta$, such that the energy bands remain deep inside the gap and the low-energy approximation holds. The Fermi wavevector determines the coupling strength between corrals. Within the approximation scheme of the low-energy model, which neglects the geometry of the quantum corrals, we expect the Fermi wavevector to be only indirectly associated to the value of the bulk material and we, therefore, set $k_{\rm F}a\sim1$. This guarantees a sufficiently strong corral hybridization for achieving non-trivial phases. Also, we choose finite and small values for the renormalized SOC strength, $\tilde{\lambda}$, throughout the plots to demonstrate non-trivial phases. The coherence length calculated at the middle of the energy gap, $\xi=\sqrt{\lambda^2+u^2_F}/\Delta$, where $u_F=k_F/m$ is the bare Fermi velocity, physically determines an effective cutoff length $N_{\rm cutoff}$ for the couplings in \cref{eq:MatrixElements2D} and distinguishes between the dilute and dense lattice limits, depending on the material/parameter choice.

First, we place $N$ d-levels in a chain geometry, separated by a lattice constant $a$, see \cref{Fig.:Schematics}(b). We consider periodic boundary conditions and the long chain limit $N\gg 1$ to plot the elements in \cref{eq:MatrixElements2D} in Fourier space, see \cref{Fig.:Chains-Effective Parameters b}. We resolve the long-range hopping and spin-flip parameters, assuming a vanishing on-site energy $E_{\rm R}=0$, see \cref{Fig.:Chains-Effective Parameters a}. Yet, $h^N(p)$ can be shifted by a non-vanishing  $E_{\rm R}$. Additionally, we demonstrate the effective singlet, $\Delta^S(p)$, and triplet, $\Delta^T(p)$, superconducting order parameters, see \cref{Fig.:Chains-Effective Parameters b}. We notice the flattening of $\Delta^{S,T}(p)$ for momenta greater than the maximum of the Fermi wavevectors of the two helicity bands, i.e., $p>{\rm max}\left(k_{F_{+}},k_{F_{-}}\right)$. This flattening is not numerically exact due to the finite Fourier transform cutoff $N_{\rm cutoff}$, see inset in \cref{Fig.:Chains-Effective Parameters b}. Furthermore, we observe characteristic cusps in the Brillouin Zone that form around the momenta $p= k_{F_\pm}$. We explain these effects by analytically computing the continuous Fourier transform of the singlet and triplet order parameters in the dense chain limit, both of which approach infinity at $p=k_{F_\pm}$, due to the $1/ \sqrt{k^2_F-p^2}$ factor, and vanish for $p>{\rm max}\left(k_{F_{+}},k_{F_{-}}\right)$ due to the combination of Heaviside $\Theta$ functions that enter the exact expressions, see \cref{Appendix:DenseChain}. The finite cusps and flattenings transcend the dense limit and survive for finite values of $k_{\rm F}a$. The singlet parameter includes both on-site and long-range elements. The observed cancellation of $\Delta^S(p)$ is lifted as $k_{\rm F}a$ increases. This is because in the ultra dilute limit $k_{\rm F}a\gg 1$, the on-site singlet superconducting pairing dominates and the system, thus, becomes topologically trivial. As shown in \cref{Fig.:Chains-Effective Parameters b}, the singlet pairing is larger than the triplet pairing in the Brillouin zone (apart from the small oscillations due to the finite cutoff). The dominance of the singlet parameter places our system in the regime where the no-go theorem of Ref.~\cite{PhysRevB.94.161110} applies. The theorem states that any non-interacting electronic system proximity-coupled to a parent superconductor is locked in the trivial phase of class DIII whenever the matrix product $\Delta(\textbf{p})\,\mathcal{T}_{\rm sc}$ is positive semi-definite (PSD) at every momentum, where $\Delta(\textbf{p})$ is the parent's superconducting pairing matrix and $\mathcal{T}_{\rm sc}$ is the unitary representation of time-reversal acting on the same internal indices. For the four-band effective Hamiltonian in \cref{eq:momentumH}, the PSD condition reduces to the inequality $|\Delta^{S}(p)|\geq|\Delta^{T}(p)|$ for every momentum $p$ (see \cref{Appendix:PSDcondition}), since the eigenvalues of $\Delta(p)\,\mathcal{T}_{\rm sc}$ are $|\Delta^{S}(p)|\pm|\Delta^{T}(p)|$. Inspection of \cref{Fig.:Chains-Effective Parameters c} shows that at $U=0$ this inequality is satisfied throughout the Brillouin zone, so the system is rigorously locked in the trivial phase. Circumventing this trivial fate requires a mechanism that drives some momentum out of the PSD region, which we now show is provided by the on-site repulsion. 

In experimental setups~\cite{Schneider2023,PhysRevB.110.L100505,Ton2025}, the confinement of the surface state in quantum corrals induces, in principle, weak repulsive particle-particle interactions, which we have so far neglected. Such interactions, effectively described by the term $U d^{\dagger}_{\uparrow,j}d_{\uparrow,j} d^{\dagger}_{\downarrow,j}d_{\downarrow,j}$ with $U>0$ in \cref{eq:Hamiltonian}, influence the singlet-triplet pairing competition. To quantify this, we consider a mean-field decoupling of the interaction term in the superconductivity channel and the mean-field parameter $\delta:=\langle d_{\uparrow,j} d_{\downarrow,j}\rangle$, see \cref{Appendix:CoulombInteraction}. We find that the on-site singlet order parameter gets renormalized $\Delta^{S,R}_{i,i}=-\Gamma+\tilde{U}$ with $\tilde{U}=-U \delta$, while the long-range singlet and all triplet pairings remain unaffected. By numerical evaluation of the mean-field parameter $\delta_{\rm R}$ which minimizes the free-energy, $\partial \mathcal{F}/\partial\delta|_{\delta=\delta_{\rm R}}=0$, in the first iteration of the self-consistency equation, we find that the singlet order parameter is suppressed, i.e., $\delta_{\rm R}<0$ or $\tilde{U}>0$, in the presence of repulsive interactions, see \cref{Appendix:CoulombInteraction} for details. Qualitatively, this suppression favors non-trivial topological phases, since the condition in \cref{eq:Condition1D} is more readily satisfied. As a consequence, $|\Delta^{S}(p)|<|\Delta^{T}(p)|$ in a finite momentum interval (see \cref{Fig.:Chains-Effective Parameters c}), so the PSD prerequisite of the no-go theorem fails and a non-trivial DIII phase becomes possible --- one of the key ideas of the present manuscript. We emphasize, however, that violating the PSD condition is necessary but not sufficient for non-trivial topology: it lifts the trivial-phase lock without guaranteeing that the system is topologically non-trivial. Determining whether a non-trivial DIII phase is realised requires explicit computation of the $\mathbb{Z}_2$ invariant $W^{1D}$, which we perform below.

\begin{figure*}[htbp]\label{Fig:1D Chains}
\begin{subfigure}[b]{0.37\textwidth}
    \centering
    \caption[]{}
    \label{Fig.:TopologicalInvariant1D-U}
\includegraphics[width=\textwidth]{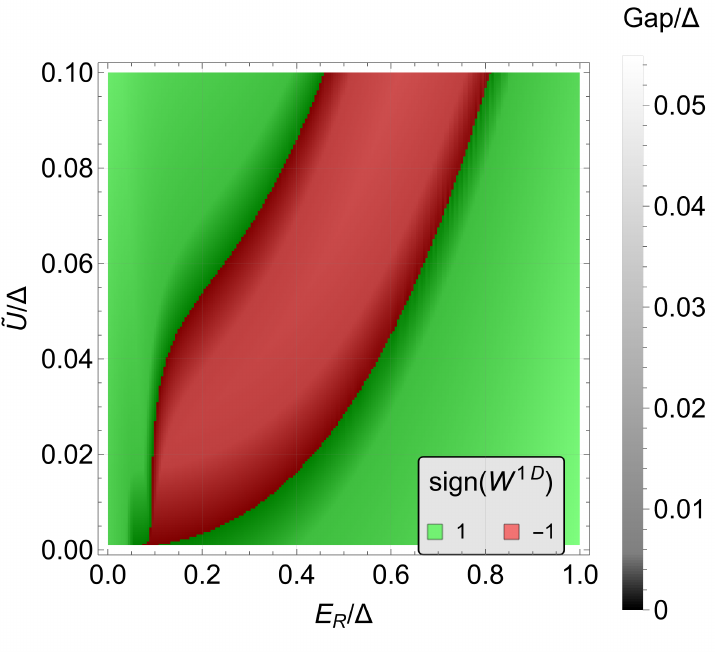}
    \end{subfigure}
    \hfill
\begin{subfigure}[b]{0.301\textwidth}
    \centering
    \caption[]{}
    \label{Fig.:TopologicalInvariant1D-kF}
\includegraphics[width=\textwidth]{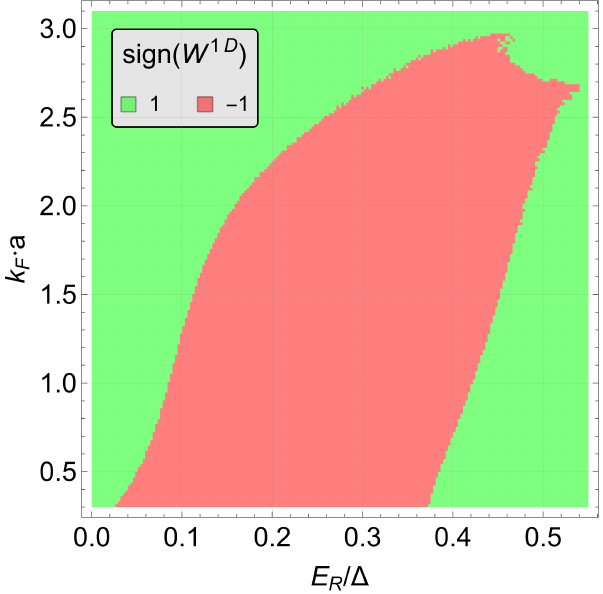}
    \end{subfigure}
    \hfill
    \begin{subfigure}[b]{0.309\textwidth}
    \centering
    \caption[]{}
    \label{Fig.:TopologicalInvariant1D-lambda}
\includegraphics[width=\textwidth]{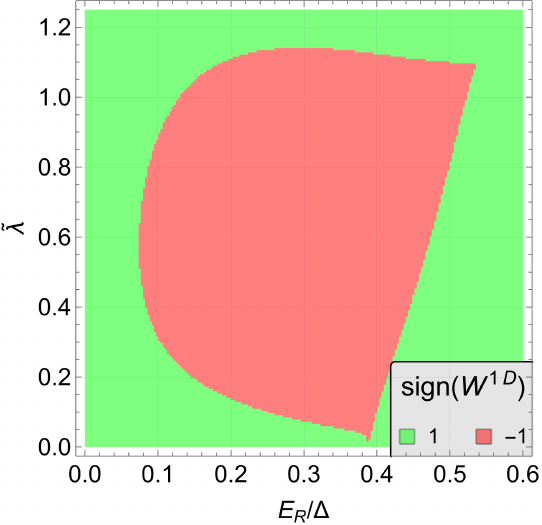}
    \end{subfigure}
    \begin{subfigure}[t]{0.5\textwidth}
    \centering
    \caption[]{}
\includegraphics[width=\textwidth]{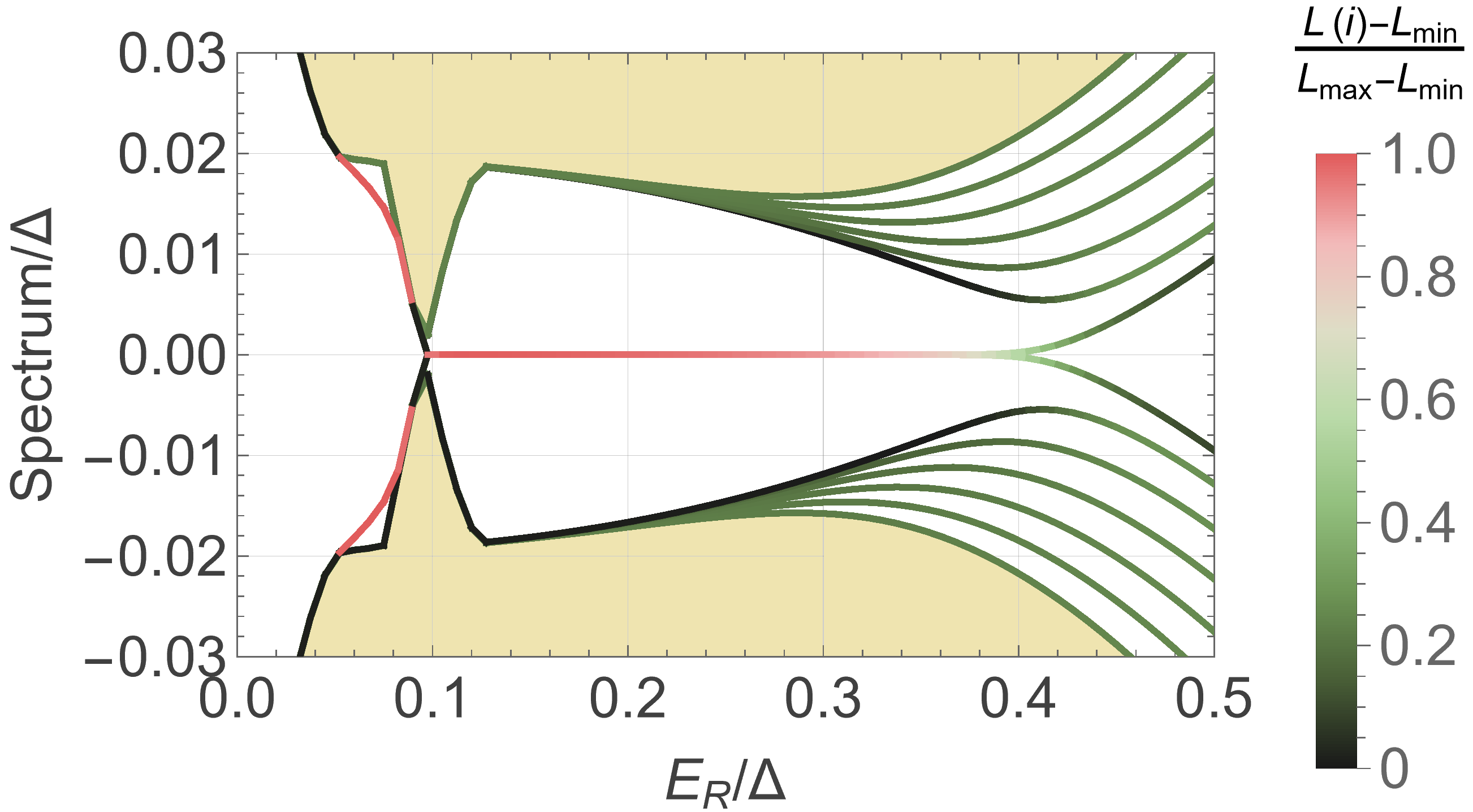}
      \label{Fig:1D Real-Space Spectrum}
    \end{subfigure}
    \hspace{0.3cm}
     \begin{subfigure}[t]{0.45\textwidth}
    \centering
    \caption[]{}
\includegraphics[width=\textwidth]{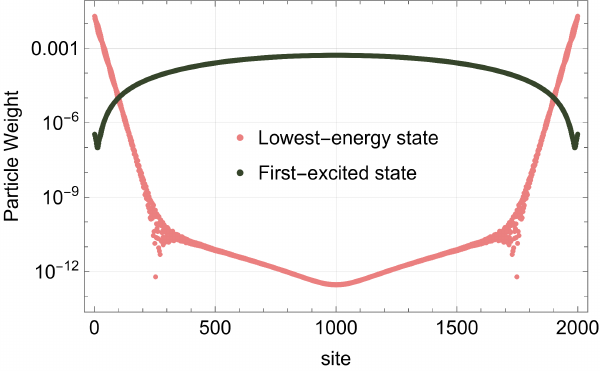}
      \label{Fig:1D Real-Space Wavefunction}
    \end{subfigure}
    \caption{Non-trivial topological phases and Majorana Kramers pairs in a one-dimensional chain with a finite repulsive interaction $\tilde{U}$. The $\mathbf{Z}_2$ topological invariant, $W^{1D}$, identifies trivial ($1$) and non-trivial ($-1$) phases, (a) depending on the repulsion interaction $\tilde{U}/\Delta$ and the on-site energy $E_{\rm R}/\Delta$, (b) depending on the on-site energy $E_{\rm R}/\Delta$ and the Fermi wavevector $k_{\rm F}a$, (c) depending on the on-site energy $E_{\rm R}/\Delta$ and the SOC $\tilde{\lambda}$. (d) Real-space spectrum (lowest energy eigenvalues) as a function of the bare energy $E_{\rm R}$ for a chain-length of $N=2000$. The yellow shading indicates higher energy states that we neglect. The color code uses the localization measure $L(i)$, see \cref{eq:LocalizationMeasure}, that demonstrates the real-space edge localization of each state with $L_{\rm min},L_{\rm max}$ being the minimum and maximum values, respectively, for all data. (e) Real-space mapping of the particle component of the wavefunction of the lowest energy (pink) and first excited (green) eigenstates along the chain at $E_{\rm R}=0.2 \Delta$ (white dot in (a)), demonstrated in logarithmic scale. Common parameter values:  $m=10^{-2}  (a^2 \Delta)^{-1}$, $\abs{V}=5\Delta$. Panel (a) $k_{\rm F}a=1.2$, $\tilde{\lambda}\equiv\lambda/u_{\rm F}=1/3$. Panel (b): $\tilde{\lambda}=1/3$, $\tilde{U}=0.02\Delta$. Panel (c):  $k_{\rm F}a=1.2$, $\tilde{U}=0.02\Delta$. Panels (d) and (e):  $k_{\rm F}a=1.2$, $\tilde{\lambda}=1/3$, $\tilde{U}=0.02\Delta$.}
    \label{Fig.:1DTop}
\end{figure*}

We demonstrate the real-space spectrum for a chain of $N$ sites as a function of the on-site energy $E_{\rm R}$, see \cref{Fig:1D Real-Space Spectrum}. Notably, there is a parameter region with the lowest eigenenergy being almost zero signifying the topologically non-trivial phase which hosts a Kramers pair of Majorana zero modes in each chain end. The zero energy eigenvalues are lifted when the spectral gap closes, leading to a topological phase transition to the trivial phase. We highlight that all eigenvalues are doubly degenerate due to time-reversal-symmetry. To quantify the real-space localization of the eigenmodes we employ the localization measure (color scale in \cref{Fig:1D Real-Space Spectrum})
\begin{equation}\label{eq:LocalizationMeasure}
L(i)=\sum_{j=1}^{j=N}\abs{\psi^i_j}^2(j-(N+1)/2)^2/((N+1)/2-1)^2,
\end{equation} where $\psi^i_j$ is the wavefunction of the  $i^{\rm th}$ eigenstate at site j. This localization measure approaches $L\approx1/3$ for a uniformly distributed state and $L\approx1$ for an edge localized state in the large-$N$ limit. Additionally, we resolve the particle part of the wavefunction of the zero-energy mode in logarithmic scale and compare it to the one of the first excited state, see \cref{Fig:1D Real-Space Wavefunction}. Evidently, the wavefunction of the zero-energy mode is exponentially localized near the boundaries of the chain by the induced coherence length, $\xi$, that enters in the matrix elements in \cref{eq:MatrixElements2D}, while the wavefunction of the first excited state is spread along the chain.

To confirm the topological origin of the zero-energy modes, we calculate the topological invariant for the effective Hamiltonian in momentum space in \cref{eq:momentumH}. For the construction of the topological invariant, we follow the approach of~\cite{PhysRevB.88.134523} and numerically compute the determinant of the projection of the Kato propagator $K$, along the path $p$ from $p=0$ to $p=\pi$, on an arbitrary choice of Bloch eigenfunctions $\det(K)$. For the construction of the topological invariant $W^{1D}$, we use $W^{1D}=\det(K) \dfrac{{\rm Pf}(\theta_{0})}{{\rm Pf}(\theta_{\pi})}$, where $\theta_p$ is the matrix representation of the time-reversal operator on the occupied Bloch eigenvectors at the time-reversal-invariant momenta, $p \in \{0,\pi\}$, and ${\rm Pf}$ denotes the Pfaffian. The determinant can take two values $\det(K)=\pm 1$ in the limit  of taking infinite steps in the partition of the path $p$ from $0\rightarrow\pi$. Therefore, the topological invariant is $W^{1D}=1$ for a trivial and $W^{1D}=-1$ for a non-trivial phase. For practical purposes, we compute the ${\rm sign}(W^{1D})$, see \cref{Fig.:1DTop}, to avoid non-quantized values that numerically emerge due to the finite discretization of the momentum interval for the Kato propagator. We show that a small repulsive interaction, $\tilde{U}>0$, is essential for opening non-trivial topological phases whose boundaries coincide with the band-gap closings, see \cref{Fig.:TopologicalInvariant1D-U}. Note that we restrict the range of our computation to small values of the repulsive interaction, $0<\tilde{U}<0.1\Delta$, where the mean-field treatment of \cref{eq:Hamiltonian} is justifiable. We demonstrate the dependence of the topological invariant on the Fermi wavevector, $k_{\rm F}$, and the on-site energy $E_{\rm R}/\Delta$, see ~\cref{Fig.:TopologicalInvariant1D-kF}. The non-trivial phase disappears above a critical value of the Fermi wavevector. Indeed, in the ultra-dilute limit $k_{\rm F} a\gg 1$, the on-site singlet superconductivity dominates and the chain is topologically trivial since the conditions in \cref{eq:Condition1D} can not be satisfied. Note that the boundary of the phase diagram for $k_{\rm F}a=1.2$ in \cref{Fig.:TopologicalInvariant1D-kF} agrees with the zero-energy crossings, shown in the finite system in \cref{Fig:1D Real-Space Spectrum}. Moreover, we study the dependence of the topological invariant, $W^{1D}$, on the renormalized SOC strength $\tilde{\lambda}$ and on-site energy $E_{\rm R}$, see \cref{Fig.:TopologicalInvariant1D-lambda}. Even in the limit of weak SOC, $\tilde{\lambda}\rightarrow0$, the system becomes non-trivial with a fine-tuning of $E_{\rm R}$. The case $\tilde{\lambda}=0$ is always trivial because the triplet order parameter vanishes. In the opposite limit of large SOC $\tilde{\lambda}=\lambda m/k_{\rm F}\gg1$, the system becomes trivial. In this limit, the coupling strength $\Gamma_{-}$ of one of the helicity bands becomes very small, see \cref{eq:supprenormalized}. In such cases, the band is pushed away from zero energy, making it impossible to experience a topological phase transition for any value of the on-site energy $E_{\rm R}$. Thus, we find only trivial phases for large values of $\tilde{\lambda}$ in \cref{Fig.:TopologicalInvariant1D-lambda}. This is in contrast to typical topological phase diagrams of YSR bands~\cite{PhysRevB.91.064505}, which demonstrate alternations between trivial and non-trivial phases when the SOC strength is increased to larger values. Last, as discussed previously in the text, we note that the other parameters of the system may change the phase boundaries in the topological phase diagrams (shown in \cref{Fig.:TopologicalInvariant1D-kF,Fig.:TopologicalInvariant1D-lambda}) but do not let the non-trivial phases completely disappear.

\begin{figure}[htbp]
    \centering
    \label{}
\includegraphics[width=0.5\textwidth]{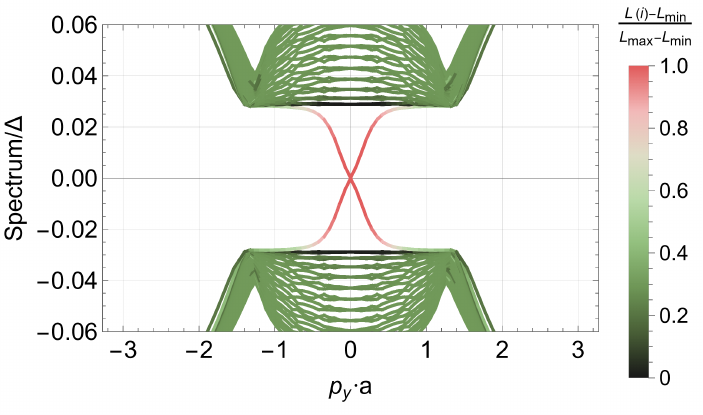}
    \caption{Localization of edge modes in two-dimensional lattices in a cylindrical geometry with finite repulsive interaction $\tilde{U}=0.05\Delta$. States are colored according to their edge localization, using the localization measure, $L(i)$, in \cref{eq:LocalizationMeasure}. (adopted parameter values: $m=10^{-2}  (a^2 \Delta)^{-1}$, $\abs{V}=5\Delta$, $\tilde{\lambda}=1/3$, $k_{\rm F}a=0.1$, $E_{\rm R}=0.1\Delta$, $N_{\rm x}=90$.)}
    \label{Fig.:2DGeometry}
\end{figure}

We next extend our study to two-dimensional square lattices with $N_{\rm x}, N_{\rm y}$ lattice sites in the $x,y$ directions, respectively, see \cref{Fig.:Schematics}(a). In this geometry, we find helical Majorana edge modes, in contrast to the end states that appear in one-dimensional chains. To demonstrate boundary modes, we consider a cylindrical geometry by taking periodic boundary conditions in the y-direction, in which case the momentum $p_{\rm y}$ is a good quantum number, and the limit $N_{\rm y}\to \infty$. In this setup, in the topologically non-trivial phase, we find time-reversal-symmetric pairs of dispersive edge modes which are localized near the open boundaries and higher energy modes which are extended throughout the bulk, see \cref{Fig.:2DGeometry}. A broader study of the topological phases in 2D lattices requires the calculation of the relevant topological invariant in the symmetry class DIII by taking two 1D cuts (for $p_y=0$ and $p_y=\pi$) in the Brillouin zone and multiplying the calculated 1D invariants in these cuts, $W^{2D}=W_{p_y=0}^{1D} \times W_{p_y=\pi}^{1D}$~\cite{PhysRevB.94.161110}. In general, we find that the 2D topological phase diagrams differ from the 1D. Specifically, the 2D equivalents of the diagrams in \cref{Fig.:TopologicalInvariant1D-kF,Fig.:TopologicalInvariant1D-lambda} demonstrate extended non-trivial phases,  see \cref{Appendix:TopologicalInvariant2D} for details. Notably, the main characteristic features for triviality for large $k_{\rm F}a$ and the non-trivial region vanishing at $\tilde{\lambda}=0$ remain. For topological phase diagramms shown in \cref{Fig:2DPhaseDiagramm}, a nearest-neighbor cutoff was considered due to computation limitations. Once the physical long-range couplings are restored, small-repulsive interactions $\tilde{U}>0$ are still required for generating non-trivial phases, similarly to the one-dimensional chains, and lead to quantitatively different topological phase diagrams. For fully periodic boundary conditions, the triplet superconducting pairing in \cref{eq:momentumH} can be simplified in the continuum limit, i.e., $|\textbf{p}|a\ll1$,
\begin{equation}
\begin{split}
     &\hat{\Delta}^{T}=\Vec{d}(\textbf{p})\Vec{\sigma}, \ \Vec{d}(\textbf{p})=ig(p) (p_x,p_y,0),\ {\rm where}\\
     &g(p)= -\sum_{R\neq0} J_1(pR) \Im(w^{o}(R))/p,
     \end{split}
\end{equation}
where $J_1$ is the Bessel function, $g(p)\in \mathbb{R}$, and the superconducting pairing is helical~\cite{PhysRevB.111.144508}. This is in contrast to the chiral triplet pairing that governs Shiba systems due to time-reversal-symmetry breaking~\cite{Li2016}.
\section{CONCLUSIONS}
 On a superconducting substrate, MS bands can be engineered by constructing arrays of adatom quantum corrals. We find that in the presence of small repulsive interactions inside the corrals, the triplet pairing can dominate over the singlet pairing leading to non-trivial topological phases in class DIII. We emphasize that such type of effective superconducting correlations is fundamentally different to the case of Shiba bands, where only effective chiral triplet pairings are present. This leads to distinct helical Majorana boundary modes in two-dimensional MS lattices and Kramers degenerate Majorana zero modes in one-dimensional chains. Interestingly, we find that the superconducting pairings become flat for a finite momenta range when long-range couplings between MSSs are not negligible. Let us address the experimental relevance of the parameters used in the model. While the effective electron mass, $m$ and spin orbit coupling $\tilde{\lambda}$ correspond to the experimental range~\cite{Schneider2023,PhysRevB.110.L100505}, we chose hybridization strengths $\Gamma$ and effective corral distances $k_{\rm F} a$ that are up to an order of magnitude smaller. First, we chose a $\Gamma \sim 0.1 \Delta$ to remain within the validity regime of the low-energy theory. Larger values of $\Gamma$ do not change the qualitative behavior of our model, but would merely distort the dispersion relation further away from the Fermi level. Topological phase transitions would still be captured faithfully because they happen at the Fermi level. Second, in an experiment, the parameter $k_{\rm F} a$ mostly describes the hybridization strength between corral modes instead of the actual Fermi momentum. An estimate of this parameter requires material-specific first-principle calculations that consider the corrals' specific geometry, which exceeds the scope of our current study and links the effective parameters in our model to the ones of the experimental configuration. A promising setup for increasing the nearest-neighbor MSSs hybridization would entail removing adatoms from the boundaries of neighboring corrals, allowing the respective MS states to hybridize sufficiently, which has been realized in elliptical and square Cu(111) corrals in the normal state~\cite{doi:10.1021/acsnano.2c00467,10.21468/SciPostPhys.9.6.085}. In the absence of superconductivity, multiple-corral setups have shown long-range hybridization patterns when placed in one and two-dimensional geometries, forming artificial molecular structures~\cite{doi:10.1126/science.adf2685}. From the typical candidate materials (Ag, Cu, Au) that host Shockley surface states on their (111) surface, Au is most promising due to the large spin-orbit coupling at its atomic levels~\cite{PhysRevLett.98.186807}. For Au, we consider $u_{F}=1.4\cdot 10^6 \text{m/s}$~\cite{Ashcroft76} and $\lambda=0.33 \text{eV\AA}$~\cite{PhysRevLett.77.3419}, leading to $\tilde{\lambda}=\lambda/(\hbar u_{\rm F})=0.04$, which is sufficiently strong to achieve non-trivial phases without exhaustive precision for tuning the chemical potential $E_{\rm R}/\Delta$. It is worth mentioning that metals with more complex electronic structures than noble metals, such as Bi and its (111) surface, are promising for experimenting with our proposal due to the possibility for epitaxial growth and large Fermi wavelength~\cite{HATTAB20088227}, despite not having a clear Shockley type of surface state.
 
Our results inspire future experiments for constructing lattices of quantum corrals and varying their mutual couplings. The construction of quantum corrals with tunable sizes offers experimental control on the corrals' on-site energies which control the topological invariant of the setup. Such lattices can host non-trivial phases and boundary modes in class DIII beyond the recently studied magnetic adatom lattices in class D. The MSSs and, thus, the boundary modes are spread within corrals of dozens of ${\rm nm}^2$ surface area~\cite{Schneider2023,PhysRevB.110.L100505,Ton2025} which enables experimental detection due to their extended real-space signal. Fabricating corral lattices would involve the lateral manipulation of potentially thousands of adatoms. Yet, this could be achieved using state-of-the-art STM manipulation techniques~\cite{Gomes2012,doi:10.1021/acsnano.2c00467,doi:10.1126/sciadv.aar5251}, or developing machine-learning-supported construction. Despite the distinct microscopic modeling of alternative platforms that host MSSs, e.g., the transmon qubit Josephson junctions~\cite{PhysRevLett.124.246802,PhysRevLett.124.246803}, our study inspires research adapted to these systems. Additionally, we envision tunable control of the edges modes hybridization by introducing weak violations of time-reversal-symmetry, e.g., with an introduction of small magnetic fields. We propose synthesizing extended lattices of magnetic adatom-quantum corral composites~\cite{Ton2025} where Kramers degeneracy is broken and topological phases beyond class DIII can be generated.
\begin{acknowledgments}
\thore{We thank Oladunjoye A. Awoga, Jens Wiebe, Levente Rózsa, and Pascal Simon for helpful
discussions. I.I. and T.P. acknowledge funding of the Cluster of Excellence `Advanced Imaging of Matter' (EXC 2056 - project ID 390715994) of the Deutsche Forschungsgemeinschaft (DFG), and T.P. acknowledges support from the European Union (ERC, QUANTWIST, project number 101039098). C.-K. C. was
supported by Japan Science and Technology Agency
(JST) as part of Adopting Sustainable Partnerships for
Innovative Research Ecosystem, Grant No. JPMJAP2318,
and by JST Presto Grant No. JPMJPR2357.} 
\end{acknowledgments}

\appendix
\renewcommand{\thefigure}{A\arabic{figure}}

\section{DERIVATION OF THE EFFECTIVE HAMILTONIAN}\label{Appendix:Derivation}
To deal with the SOC term in \cref{eq:Hamiltonian}, we transform our bulk Hamiltonian $H_{\rm SC}$ to the helicity basis
\begin{equation}\label{eq:suppbasis}
\begin{split}
c_{\textbf{k}_+}=&\dfrac{1}{\sqrt{2}}\left(ie^{-i\theta(\textbf{k})}c_{\textbf{k} \downarrow}+c_{\textbf{k} \uparrow}\right),
\\
    c_{\textbf{k}_-}=&\dfrac{1}{\sqrt{2}}\left(c_{\textbf{k} \downarrow}+ie^{i\theta(\textbf{k})}c_{\textbf{k} \uparrow}\right),
    \end{split}
    \end{equation}
    where $\theta(\textbf{k})$ is the azimuthal angle for the momentum of the $c$ electrons. The bulk Hamiltonian is rewritten 
    \begin{equation}\label{eq:suppHhbasis}
    \begin{split}
    H_{\pm}=&\sum_{\textbf{k},\pm} \epsilon_{\textbf{k}_\pm} c^{\dagger}_{\textbf{k}_\pm} c_{\textbf{k}_\pm}\\&-\sum_{\textbf{k},\pm} \Delta^*_{\textbf{k}_\pm}c^{\dagger}_{\textbf{\textbf{k}}_\pm} c^{\dagger }_{-\textbf{k}_\pm}+\Delta_{\textbf{k}_\pm}c_{-\textbf{k}_\pm}c_{\textbf{k} \pm},
    \end{split}
\end{equation}
where $\epsilon_{\textbf{k}_\pm}=\dfrac{\textbf{k}^2}{2m}-\epsilon_F\pm \lambda |\textbf{k}|$ and $\Delta_{\textbf{k}_\pm}=\Delta e^{\pm i\left(\theta(\textbf{k})+\frac{\pi}{2}\right)}$. The tunneling in \cref{eq:Hamiltonian} in the basis (\ref{eq:suppbasis}) is
\begin{equation}\label{eq:supptransH}
\begin{split}
    H_T=&\dfrac{V}{\sqrt{2}}\sum_{\textbf{k},\sigma,j} e^{i\textbf{k}\textbf{R}_j}(\left( ie^{i\theta(\textbf{k})}c^{\dagger}_{\textbf{k}_-}+ c^{\dagger}_{\textbf{k}_+}\right)d_{\uparrow,j}\\&
    +\left( c^{\dagger}_{\textbf{k}_-}+ie^{-i\theta(\textbf{k})}c^{\dagger}_{\textbf{k}_+}\right)d_{\downarrow,j} )+h.c  \ .
\end{split}
\end{equation}
We next use the Green's function equations of motion to integrate out the bulk modes and derive the effective Hamiltonian in \cref{eq:momentumH}. We define the matrix Green's function in Zubarev's notation~\cite{Zubarev} of the d-levels
\begin{equation} \label{eq:suppmatrixGF}
   \check{G}=
   \ll
   \begin{pmatrix}
         d_{\uparrow,i}\\
        d_{\downarrow,i}\\
        d^{\dagger}_{\downarrow,i}\\
        d^{\dagger}_{\uparrow,i} 
    \end{pmatrix};\begin{pmatrix}
         d^{\dagger}_{\uparrow,j} &
        d^{\dagger}_{\downarrow,j} &
        d_{\downarrow,j} &
        d_{\uparrow,j} 
    \end{pmatrix}
    \gg.
\end{equation}
In general, we can write 
\begin{equation}\label{eq:suppmatrixGFEOM}
      (E-\hat{H}_{\rm eff}(E)) \check{G}=\hat{1},
\end{equation}
where the matrix $\hat{H}_{\rm eff}(E)$ is derived in the following by the Green's function equations of motion~\cite{8857dd78fb874b8ca4eb9995a2671dfc}. The Green's functions of the d-levels are 
\begin{equation}\label{eq:suppdEOM}
\begin{split}
    (E-E_{\rm R}) G_{d_{\uparrow,m} d^{\dagger}_{\uparrow,m}}=&1+\dfrac{V^*}{\sqrt{2}} \sum_\textbf{k}   \left(-ie^{-i\theta}G_{\textbf{k} _- d^{\dagger}_{\uparrow,m}}\right)+\\&\dfrac{V^*}{\sqrt{2}} \sum_\textbf{k} G_{\textbf{k} _+d^{\dagger}_{\uparrow,m}},\\
    (E-E_{\rm R}) G_{d_{\downarrow,m} d^{\dagger}_{\downarrow,m}}=&1+\dfrac{V^*}{\sqrt{2}} \sum_\textbf{k} G_{\textbf{k} _- d^{\dagger}_{\downarrow,m}}+\\&\dfrac{V^*}{\sqrt{2}} \sum_\textbf{k}  \left( -ie^{i\theta}G_{\textbf{k} _+d^{\dagger}_{\downarrow,m}}\right).
    \end{split}
\end{equation}
We then substitute the sums of the Green's function that involve bulk modes and find 
\begin{widetext}
\begin{equation}\label{eq:suppbulkGFeom}
\begin{split}
  &(E-\epsilon_{\textbf{k}_+}) G_{\textbf{k}_+ d^{\dagger}_{\uparrow,m}} =\sum_{i}\left(\dfrac{V_i}{\sqrt{2}} G_{d_{\uparrow,i} d^{\dagger}_{\uparrow,m}}+\dfrac{iV_ie^{-i\theta} }{\sqrt{2}} G_{d_{\downarrow,i} d^{\dagger}_{\uparrow,m}}\right) -\Delta^*_+G_{-\textbf{k}^{\dagger}_{+} d^{\dagger}_{\uparrow,m}},
\\
   &(E+\epsilon_{\textbf{k}_+}) G_{-\textbf{k}^{\dagger}_+d^{\dagger}_{\uparrow}}=\sum_{i}\left(- \dfrac{iV^*_{i,-\textbf{k}}e^{i\theta} }{\sqrt{2}}  G_{d^{\dagger}_{\downarrow,i} d^{\dagger}_{\uparrow,m}}-  \dfrac{V^*_{i,-\textbf{k}}}{\sqrt{2}}  G_{d^{\dagger}_{\uparrow,i} d^{\dagger}_{\uparrow,m}}\right)
   -\Delta_+  G_{\textbf{k}_+d^{\dagger}_{\uparrow,m}},
\\
  &(E-\epsilon_{\textbf{k}_-}) G_{\textbf{k}_- d^{\dagger}_{\uparrow,m}} =\sum_i\left(\dfrac{iV_ie^{i\theta}}{\sqrt{2}}G_{d_{\uparrow,i} d^{\dagger}_{\uparrow,m}}+\dfrac{V_i}{\sqrt{2}} G_{d_{\downarrow,i} d^{\dagger}_{\uparrow,m}}\right) -\Delta^*_- G_{-\textbf{k}^{\dagger}_{-} d^{\dagger}_{\uparrow,m}},
\\
   &(E+\epsilon_{\textbf{k}_-}) G_{-\textbf{k}^{\dagger}_-d^{\dagger}_{\uparrow,m}}=\sum_{i} \left(-\dfrac{V^*_{i,-\textbf{k}}}{\sqrt{2}}  G_{d^{\dagger}_{\downarrow,i} d^{\dagger}_{\uparrow,m}}- \dfrac{iV^*_{i,-\textbf{k}}e^{-i\theta}}{\sqrt{2}}  G_{d^{\dagger}_{\uparrow,i} d^{\dagger}_{\uparrow,m}}\right)
 -\Delta_-  G_{\textbf{k}_-d^{\dagger}_{\uparrow,m}}.
   \end{split}
\end{equation}
By solving \cref{eq:suppbulkGFeom} and substituting in the first \cref{eq:suppdEOM}, we obtain
\begin{equation} \label{eq:suppFirst}
\begin{split}
   &\left(E-E_{\rm R}+\dfrac{E \Gamma}{\sqrt{\Delta^2-E^2}}\right)G_{d_{\uparrow,m}d^{\dagger}_{\uparrow,m}}=1
   -\dfrac{\Delta \Gamma }{\sqrt{\Delta^2-E^2}}G_{d^{\dagger}_{\downarrow,m} d^{\dagger}_{\uparrow,m}}+\sum_{j}G_{d_{\uparrow,j}d^{\dagger}_{\uparrow,m}}\left(S_{3,j,m}+S_{4,j,m}\right)\\ &+\sum_{j}
    G_{d_{\downarrow,j} d^{\dagger}_{\uparrow,m}}\left(S_{5,j,m}+S_{6,j,m}\right)+\sum_{j}G_{d^{\dagger}_{\downarrow,j} d^{\dagger}_{\uparrow,m}}S_{7,j,m}-\sum_{j} G_{d^{\dagger}_{\uparrow,j} d^{\dagger}_{\uparrow,m}}S_{8,j,m},
    \end{split}
\end{equation}
and, similarly, for the opposite spin species 
\begin{equation} \label{eq:suppSec}
\begin{split}
   &\left(E-E_{\rm R}+\dfrac{E \Gamma}{\sqrt{\Delta^2-E^2}}\right)G_{d_{\downarrow,m}d^{\dagger}_{\downarrow,m}}=1
   +\dfrac{\Delta \Gamma }{\sqrt{\Delta^2-E^2}}G_{d^{\dagger}_{\uparrow,m} d^{\dagger}_{\downarrow,m}}+\sum_{j}G_{d_{\downarrow,j}d^{\dagger}_{\downarrow,m}}\left(S_{3,j,m}+S_{4,j,m}\right)\\ &-\sum_{j}
    G_{d_{\uparrow,j} d^{\dagger}_{\downarrow,m}}\left(S_{5,j,m}+S_{6,j,m}\right)^*-\sum_{j}G_{d^{\dagger}_{\uparrow,j} d^{\dagger}_{\downarrow,m}}S_{7,j,m}-\sum_{j} G_{d^{\dagger}_{\downarrow,j} d^{\dagger}_{\downarrow,m}}S^*_{8,j,m},
    \end{split}
\end{equation}
with the definitions
\begin{equation} \label{eq:suppSDEF}
\begin{split}
    S_{3,j,m}=&\dfrac{\abs{V}^2}{2}\sum_{\textbf{k},\mu=\pm}\dfrac{e^{i \textbf{k} \left(\textbf{R}_j-\textbf{R}_m\right)}E}{E^2-\epsilon^2_{\textbf{k}_\mu}-\Delta^2}, \quad S_{4,j,m}=\dfrac{\abs{V}^2}{2}\sum_{\textbf{k},\mu=\pm}\dfrac{e^{i \textbf{k} \left(\textbf{R}_j-\textbf{R}_m\right)}
          \epsilon_{\textbf{k}_\mu}}{E^2-\epsilon^2_{\textbf{k}_\mu}-\Delta^2}, \\
          S_{5,j,m}=&\dfrac{\abs{V}^2 }{2}\sum_{\textbf{k},\mu=\pm}\dfrac{i \mu e^{-i\theta}e^{i \textbf{k} \left(\textbf{R}_j-\textbf{R}_m\right)}E}{E^2-\epsilon^2_{\textbf{k}_\mu}-\Delta^2}, \quad
          S_{6,j,m}=\dfrac{\abs{V}^2 }{2}\sum_{\textbf{k},\mu=\pm}\dfrac{i \mu e^{-i\theta}e^{i \textbf{k} \left(\textbf{R}_j-\textbf{R}_m\right)}
          \epsilon_{\textbf{k}_\mu}}{E^2-\epsilon^2_{\textbf{k}_\mu}-\Delta^2} ,\\
           S_{7,j,m}=& \dfrac{\abs{V}^2}{2}\sum_{\textbf{k},\mu=\pm}\dfrac{e^{i \textbf{k} \left(\textbf{R}_j-\textbf{R}_m\right)}\Delta}{E^2-\epsilon^2_{\textbf{k}_\mu}-\Delta^2}, \quad
           S_{8,j,m}= \dfrac{\abs{V}^2 }{2}\sum_{\textbf{k},\mu=\pm}\dfrac{i \mu e^{-i\theta}e^{i \textbf{k} \left(\textbf{R}_j-\textbf{R}_m\right)}
          \Delta }{E^2-\epsilon^2_{\textbf{k}_\mu}-\Delta^2}.
\end{split}
\end{equation}
\end{widetext}
Note that  $S_3,S_4,S_7$ are even and $S_5,S_6,S_8$ are odd under the exchange $i\leftrightarrow j$. Also, $S_3,S_4,S_7$ are real. Importantly, the elements $S_5,S_6,S_8$ vanish for vanishing SOC in the bulk. In the low-energy limit, $E/\Delta\rightarrow 0$, $S_3,S_5$ vanish and \cref{eq:suppmatrixGFEOM} becomes an eigenvalue equation with the effective Hamiltonian in \cref{eq:bdgH}. We find
\begin{equation}\label{eq:suppdefmatel}
\begin{split}
    h^N_{i,j}=&E_{\rm R}\delta_{i,j}+(1-\delta_{i,j})\lim_{\frac{E}{\Delta}\rightarrow0}S_{4,i,j},
    \\
    h^F_{i,j}=&(1-\delta_{i,j})\lim_{\frac{E}{\Delta}\rightarrow0}S_{6,i,j},\\
    \Delta^S_{i,j}=&-\Gamma \delta_{i,j}+(1-\delta_{i,j})\lim_{\frac{E}{\Delta}\rightarrow0}S_{7,i,j},\\
     \Delta^T_{i,j}=&-(1-\delta_{i,j})\lim_{\frac{E}{\Delta}\rightarrow0}S_{8,i,j},
\end{split}
\end{equation}
where specific formulas for these matrix elements are demonstrated in Sec.~\ref{Sec.Results} in the main text and \cref{Appendix:2DbulkSystem}, after appropriate specifications and approximations specified there.
\section{EIGENVALUE CALCULATION}\label{Appendix:eigenvaluecalculation}
To calculate the eigenvalues for the general problem, we Fourier transform the real space functions $S_{i}$, $F_i(\textbf{p})= \sum_{j}  e^{i \textbf{p} \textbf{R}_j} S_{i,0,j}$ and rewrite the complex functions $F_6=\Re{F_6}+i\Im{F_6}$ and $F_8=\Re{F_8}+i\Im{F_8}$. In the following, we work with the momentum space functions. We write
\begin{equation}
\begin{split}
H(p)=&\tau_z\sigma_0(F_4+E_{\rm R})-\tau_0\sigma_y \Im{F_6}+\tau_0\sigma_x \Re{F_6}\\&+\tau_x \sigma_z (-\Gamma+F_7)+\tau_y \sigma_x \Im{F_8}+\tau_y \sigma_y \Re{F_8}.
\end{split}
\end{equation}
We square the Hamiltonian and obtain
\begin{equation}\label{eq:FirstSquare}
\begin{split}
&H^2-(F_4+E_{\rm R})^2-(F_7-\Gamma)^2-\abs{F_6}^2- \abs{F_8}^2=\\ &-2\tau_z\sigma_y\left((F_4+E_{\rm R})\Im{F_6}+(F_7-\Gamma)\Im{F_8}\right)\\&+2\tau_z\sigma_x\left((F_4+E_{\rm R})\Re(F_6)+(F_7-\Gamma)\Re{F_8}\right)\\&+2\tau_y\sigma_0\left(\Re{F_6} \Im{F_8}-\Im{F_6} \Re{F_8}\right).
\end{split}
\end{equation}
The above is further simplified by considering the identity $\Re{F_6} \Im{F_8}-\Im{F_6} \Re{F_8}=\Im{F^*_6F_8}$. \cref{eq:FirstSquare} is squared again and projected to the energy eigenstates to give an eigenvalue equation
\begin{equation}\label{eq:suppEigGeneral}
\begin{split}
&\dfrac{1}{4}\left(E^2-(F_4+E_{\rm R})^2-(F_7-\Gamma)^2-\abs{F_6}^2- \abs{F_8}^2\right)^2=\\ &\abs{F_4+E_{\rm R}}^2\abs{F_6}^2+\abs{F_7-\Gamma}^2\abs{F_8}^2+\Im{ F_6 F^*_8}^2\\&+2(F_4+E_{\rm R})(F_7-\Gamma)\Re{ F_6 F^*_8}.
\end{split}
\end{equation}
In the 1D case, $\Re{F_6}=\Re{ F_8}=0$, and \cref{eq:suppEigGeneral} simplifies to
\begin{equation}\label{eq:suppEigGeneral1D}
\begin{split}
&E_{\pm,\pm}=\pm\sqrt{\left(F_4+E_{\rm R}\pm  \Im{F_6}\right)^2+\left(-\Gamma+F_7\pm \Im{F_8}\right)^2}\\&=
\pm\sqrt{\left(h^N(p)\pm \Im{h^F(p)}\right)^2+\left(\Delta^S(p)\mp \Im{\Delta^T(p)}\right)^2},
\end{split}
\end{equation}
where we use the relations in \cref{eq:suppdefmatel} in momentum space for the second equality.
\section{2D-BULK SYSTEM}\label{Appendix:2DbulkSystem}
The sums in \cref{eq:suppSDEF} are solved in a 2D bulk system. We linearize the dispersion of the bulk $\epsilon_{\textbf{k}_\pm}= \left(\textbf{k}^2-k^2_F \right)/2m\pm \lambda\abs{\textbf{k}}$ to get the renormalized parameters 
\begin{equation}\label{eq:supprenormalized}
\begin{split}
    k_{\pm}=&k_{F_\pm}+ \epsilon_{\textbf{k}_\pm}/\tilde{\upsilon}_F,
    k_{F_\pm}=k_{\rm F} \left( \sqrt{\tilde{\lambda}^2+1}\mp\tilde{\lambda} \right),\\  \tilde{\upsilon}_{F}=&u_F \sqrt{\tilde{\lambda}^2+1}, \tilde{\lambda}=\lambda m/k_{\rm F}, \xi=\tilde{\upsilon}_F/\sqrt{\Delta^2-E^2},
    \\\nu_{\pm}=& \dfrac{m}{2\pi} \left(1\mp \dfrac{\tilde{\lambda}}{\sqrt{\tilde{\lambda}^2+1}} \right).
    \end{split}
\end{equation}
We also consider a 2D lattice of MS impurities at positions $\textbf{R}_j=\abs{\textbf{R}_j}\left(\cos(\phi_j),\sin(\phi_j)\right)$. Using the above, we write the relevant integrals
\begin{equation}
\begin{split}
          S^{2D}_{4,j,m}=&\dfrac{|V|^2}{2}\sum_{\textbf{k},\mu=\pm}\dfrac{e^{i \abs{\textbf{k}} \abs{\textbf{R}_j-\textbf{R}_m}\cos(\theta-\phi_{j,m})}
          \epsilon_{\textbf{k}_\mu} }{E^2-\epsilon^2_{\textbf{k}_\mu}-\Delta^2}, \\
         S^{2D}_{6,j,m}=&\dfrac{|V|^2}{2}\sum_{\textbf{k},\mu=\pm}\dfrac{i\mu e^{-i\theta}e^{i \abs{\textbf{k}} \abs{\textbf{R}_j-\textbf{R}_m} \cos(\theta-\phi_{j,m})}
          \epsilon_{\textbf{k}_\mu}}{E^2-\epsilon^2_{\textbf{k}_\mu}-\Delta^2},\\
          S^{2D}_{7,j,m}=& \dfrac{\Delta |V|^2}{2}\sum_{\textbf{k},\mu=\pm}\dfrac{e^{i \abs{\textbf{k}} \abs{\textbf{R}_j-\textbf{R}_m} \cos(\theta-\phi_{j,m})}}{E^2-\epsilon^2_{\textbf{k}_\mu}-\Delta^2},\\
          S^{2D}_{8,j,m}=& \dfrac{\Delta |V|^2}{2}\sum_{\textbf{k},\mu=\pm}\dfrac{i\mu e^{-i\theta}e^{i\abs{\textbf{k}} \abs{\textbf{R}_j-\textbf{R}_m} 
           \cos(\theta-\phi_{j,m})}
          }{E^2-\epsilon^2_{\textbf{k}_\mu}-\Delta^2}.
\end{split}
\end{equation}
We define $ x^\pm_{j,m}=\left(k_{F_{\pm}}+i\xi^{-1}\right)\abs{\textbf{R}_{j}-\textbf{R}_{m}}$. In the limit $\lim_{\frac{E}{\Delta}\rightarrow0}$, we find~\cite{Li2016,PhysRevB.91.064505,Heimes_2015} 
\begin{equation} \label{eq:Integrals2DBulk}
\begin{split}
\lim_{\frac{E}{\Delta}\rightarrow0}S^{2D}_{4,j,m}=&\sum_{\mu=\pm}\dfrac{\Gamma_{\mu}}{2}\Im{J_0\left[x^{\mu}_{j,m}\right]+iH_0\left[x^{\mu}_{j,m}\right]},\\
 \lim_{\frac{E}{\Delta}\rightarrow0} S^{2D}_{6,j,m}=&e^{-i\phi_{j,m}}\sum_{\mu=\pm}\dfrac{ \mu \Gamma_{\mu}}{2}\Re{i J_1\left[x^{\mu}_{j,m}\right]+H_{-1}\left[x^{\mu}_{j,m}\right]},\\
 \lim_{\frac{E}{\Delta}\rightarrow0} S^{2D}_{7,j,m}=&-\sum_{\mu=\pm}\dfrac{\Gamma_{\mu}}{2} \Re{J_0\left[x^{\mu}_{j,m}\right]+iH_0\left[x^{\mu}_{j,m}\right]},\\
 \lim_{\frac{E}{\Delta}\rightarrow0} S^{2D}_{8,j,m}=&e^{-i\phi_{j,m}}\sum_{\mu=\pm}\dfrac{ \mu  \Gamma_{\mu}}{2}\Im{i J_1\left[x^{\mu}_{j,m}\right]+H_{-1}\left[x^{\mu}_{j,m}\right]},
\end{split}
\end{equation}
where $\Gamma_{\pm}=\pi \nu_{\pm} \abs{V}^2$ and $J_n$ and $H_n$ are the $n^\text{th}$ Bessel and Struve functions, respectively.

\section{POSITIVE SEMI-DEFINITE CONDITION FOR THE SUPERCONDUCTING PAIRING}\label{Appendix:PSDcondition}
In this Appendix, we derive the inequality $|\Delta^{S}(\textbf{p})|\geq|\Delta^{T}(\textbf{p})|$, which the no-go theorem of Ref.~\cite{PhysRevB.94.161110} imposes on the four-band effective Hamiltonian in \cref{eq:momentumH} as a condition for being locked in the trivial phase of class DIII. In the standard Nambu basis defined below \cref{eq:momentumH}, the pairing of spinors $d^{\dagger}d^{\dagger}$ enter the bilinear $\frac{1}{2}\Psi^{\dagger}H_{\rm eff}\Psi$ through the upper right $2\times 2$ block of $H_{\rm eff}(\textbf{p})$,
\begin{equation}\label{eq:Pblock}
\begin{pmatrix} \Delta^{S}(\textbf{p}) & \Delta^{T}(\textbf{p}) \\ -\Delta^{T*}(\textbf{p}) & -\Delta^{S}(\textbf{p}) \end{pmatrix},
\end{equation}
where the time-reversal symmetry of $H_{\rm eff}$ enforces $\Delta^{S}(\textbf{p})\in\mathbb{R}$. Because the hole sector of the basis lists spin in the order $(\downarrow,\uparrow)$ rather than $(\uparrow,\downarrow)$, a $\sigma_x$ permutation of the columns relates \cref{eq:Pblock} to the canonical pairing matrix $\Delta(\textbf{p})$ entering $\tfrac12\sum_{\textbf{p},\alpha\beta}\Delta_{\alpha\beta}(\textbf{p})\,d^{\dagger}_{\alpha,\textbf{p}}d^{\dagger}_{\beta,-\textbf{p}}$:
\begin{equation}\label{eq:DeltaSpin}
\Delta(\textbf{p}) \;=\; \begin{pmatrix} \Delta^{T}(\textbf{p}) & \Delta^{S}(\textbf{p}) \\ -\Delta^{S}(\textbf{p}) & -\Delta^{T*}(\textbf{p}) \end{pmatrix}.
\end{equation}
The required antisymmetry $\Delta(-\textbf{p})^{\!\top}=-\Delta(\textbf{p})$ follows from the parities $\Delta^{S}(-\textbf{p})=\Delta^{S}(\textbf{p})$ and $\Delta^{T}(-\textbf{p})=-\Delta^{T}(\textbf{p})$. The antiunitary time-reversal operator $\mathcal{T}=\mathcal{U}_{\mathcal{T}}K$ is invariant under $\mathcal{U}_{\mathcal{T}}\to e^{i\phi}\mathcal{U}_{\mathcal{T}}$ for any $\phi\in\mathbb{R}$; the unitary representation $\mathcal{T}_{\rm sc}$ acting on the parent superconductor's spin index therefore carries one free overall phase. Following Ref.~\cite{PhysRevB.94.161110}, we exercise this freedom to render $\Delta(\textbf{p})\,\mathcal{T}_{\rm sc}$ positive semi-definite. Writing
\begin{equation}\label{eq:DeltaSphase}
\Delta^{S}(\textbf{p}) \;=\; |\Delta^{S}(\textbf{p})|\,e^{i\theta},
\end{equation}
with $\theta\in\{0,\pi\}$ a constant phase across the Brillouin zone (and $\theta=\pi$ in our regime, since $\Delta^{S}(\textbf{p})<0$ throughout, see \cref{Fig.:Chains-Effective Parameters b}), we choose
\begin{equation}\label{eq:Tscchoice}
\mathcal{T}_{\rm sc} \;=\; -i\,e^{-i\theta}\sigma_y,
\end{equation}
which reduces to the standard $\mathcal{T}_{\rm sc}=i\sigma_y$ when $\theta=\pi$. Substituting \cref{eq:DeltaSphase,eq:Tscchoice} into \cref{eq:DeltaSpin} and computing the matrix product entry by entry yields
\begin{equation}\label{eq:DeltaTsc}
\Delta(\textbf{p})\,\mathcal{T}_{\rm sc} \;=\; \begin{pmatrix} |\Delta^{S}(\textbf{p})| & -e^{-i\theta}\,\Delta^{T}(\textbf{p}) \\ -e^{-i\theta}\,\Delta^{T*}(\textbf{p}) & |\Delta^{S}(\textbf{p})| \end{pmatrix},
\end{equation}
which is Hermitian for $\theta\in\{0,\pi\}$. The structure in \cref{eq:DeltaTsc} yields the spectrum
\begin{equation}\label{eq:PSDeigenvalues}
\mathrm{eig}\bigl[\Delta(\textbf{p})\,\mathcal{T}_{\rm sc}\bigr] \;=\; |\Delta^{S}(\textbf{p})| \;\pm\; |\Delta^{T}(\textbf{p})|,
\end{equation}
since the phase $e^{-i\theta}$ rotates the in-plane $\bm{d}$-vector but does not alter its magnitude. Both eigenvalues are non-negative if and only if
\begin{equation}\label{eq:PSDcondition}
\Delta(\textbf{p})\,\mathcal{T}_{\rm sc} \;\succeq\; 0 \;\;\leftrightarrow\;\; |\Delta^{S}(\textbf{p})| \;\geq\; |\Delta^{T}(\textbf{p})| \quad\text{for every } \textbf{p},
\end{equation}
which is the inequality cited in the main text.

\twocolumngrid
\begin{figure*}[hbt!]
    \begin{subfigure}{0.55\textwidth}
    \centering
    \caption[]{}
    \label{Fig.:SelfConsistentLines}
\includegraphics[width=\textwidth]{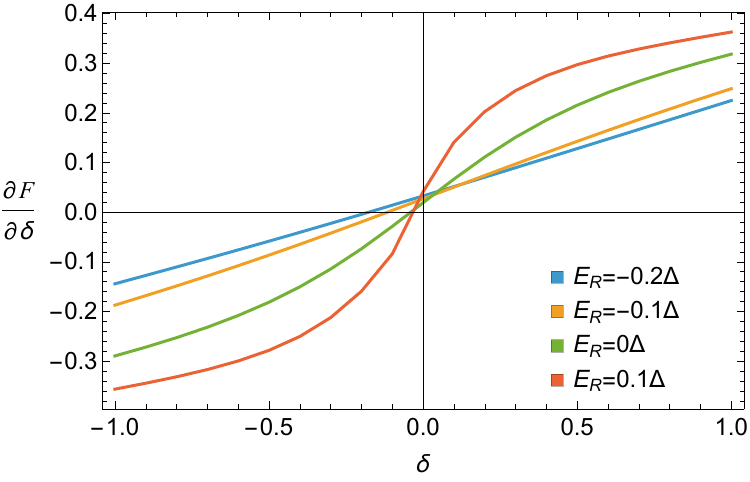}
    \end{subfigure}
    \hfill
    \begin{subfigure}{0.41\textwidth}
    \centering
    \caption[]{}
    \label{Fig.:SelfConsistentDensity}
\includegraphics[width=\textwidth]{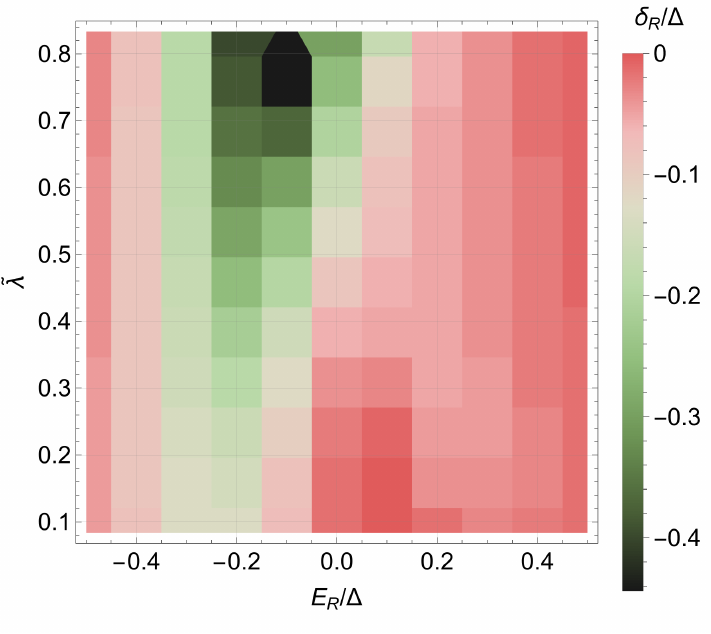}
    \end{subfigure}
    \caption{Suppression of singlet superconductivity in the presence of repulsive on-site interactions, U. (a) Free energy derivative with respect to the mean-field parameter, $\partial\mathcal{F}/\partial \delta$, as a function of $\delta$ for various $E_{\rm R}/\Delta$ around the Fermi level. (b) Mean-field parameter, $\delta_{\rm R}$, that minimizes the free energy as a function of the renormalized spin-orbit coupling strength $\tilde{\lambda}$ and  $E_{\rm R}/\Delta$. (parameters: $U=0.1\Delta$, $\beta=10$, $k_{\rm F}a=1.2$, $m=10^{-2}  (a^2 \Delta)^{-1}$, $\abs{V}=5\Delta$, a cutoff $N=300$ and, $\tilde{\lambda}=0.25$ for (a).)}
    \label{Fig.:SelfConsistent}
\end{figure*}
\section{EFFECTS OF COULOMB INTERACTION}\label{Appendix:CoulombInteraction}
We consider small repulsive interactions $U\ll\Gamma$ and $U>0$, introduced in \cref{eq:Hamiltonian}. In this limit, a mean-field approximation of the interaction term is considered~\cite{Coleman_2015}
\begin{equation}
\begin{split}
U d^{\dagger}_{\uparrow,j}d_{\uparrow,j} d^{\dagger}_{\downarrow,j}d_{\downarrow,j} \rightarrow -U\langle d^{\dagger}_{\uparrow,j}d^{\dagger}_{\downarrow,j}\rangle d_{\uparrow,j} d_{\downarrow,j}\\-U\langle d_{\uparrow,j} d_{\downarrow,j}\rangle  d^{\dagger}_{\uparrow,j}d^{\dagger}_{\downarrow,j}+U\langle d_{\uparrow,j} d_{\downarrow,j}\rangle  \langle d^{\dagger}_{\uparrow,j}d^{\dagger}_{\downarrow,j}\rangle.
\end{split}
\end{equation}
We define $\delta:=\langle d_{\uparrow,j} d_{\downarrow,j}\rangle$, which needs to be treated self-consistently and assumed to be real. Since the interaction term does not involve bulk modes, the terms can be directly transferred to the effective Hamiltonian
\begin{equation}
\begin{split}
    &H_{\rm U}=\sum_{i,j}( h^{N}_{i,j}d^{\dagger}_{\uparrow,j}d_{\uparrow,i}+h^{N}_{i,j}d^{\dagger}_{\downarrow,j}d_{\downarrow,i}+h^{F}_{i,j} d^{\dagger}_{\uparrow,j}d_{\downarrow,i}\\&+\Delta^{S,R}_{i,j} d^{\dagger}_{\uparrow,j}d^{\dagger}_{\downarrow,i}+ \Delta^{T}_{i,j}d^{\dagger}_{\uparrow,j}d^{\dagger}_{\uparrow,i}+\Delta^{T*}_{i,j}d^{\dagger}_{\downarrow,j}d^{\dagger}_{\downarrow,i})+U\delta^2+h.c,
\end{split}
\end{equation}
where the renormalized singlet superconductivity term reads $\Delta^{S,R}_{i,j}=\Delta^{S}_{i,j}-U\delta \delta^K_{i,j}$, where $\delta^K_{i,j}$ is the Kronecker delta function. The mean-field parameter $\delta$ needs to be treated self-consistently. We write the partition function of the free theory
\begin{equation}
\mathcal{Z}=e^{-\beta U\delta^2}\prod_{\textbf{k},m}\left(1+e^{-\beta E_{m}(\textbf{k})}\right),
\end{equation}
where the product is over the occupied bands. The free energy is
\begin{equation}
    \mathcal{F}=- T\int d\textbf{k}\sum_{m} \ln\left(1+e^{-\beta E_{m}(\textbf{k})}\right)+U\delta^2.
    \end{equation}
The minimization of the free energy reads
\begin{equation} \label{eq:selfconsistent}
    \dfrac{\partial\mathcal{F}}{\partial \delta}\bigg|_{\delta=\delta_R}= 2U\delta_R+\sum_{m}\int d\textbf{k}\   n_F(E_m) \dfrac{\partial E_m}{\partial\delta}\bigg|_{\delta=\delta_R}=0,
\end{equation}
where $n_F$ is the Fermi distribution. In our case, the index $m$ takes distinct values for the two helicity bands of our system. In \cref{Fig.:SelfConsistentLines}, we show the derivative of the free energy with respect to the order parameter $\partial\mathcal{F}/\partial \delta$ for different values of $E_{\rm R}/\Delta$. The sign of the root of this derivative, $\delta_{\rm R}$, determines the qualitative influence of the Coulomb interaction on the on-site superconductivity strength. The mean-field \cref{eq:selfconsistent} is approximated in the first iteration of self-consistency. If $\delta_{\rm R}> 0$ it is enhanced, while  for $\delta_{\rm R}<0$ it is suppressed. To this end, we show in \cref{Fig.:SelfConsistentDensity} that $\delta_{\rm R}$ is typically negative for a wide range of parameters when considering a large real-space cutoff. Thus, we conclude that repulsive Coulomb interactions suppress the singlet superconductivity, allowing the triplet part to dominate in parts of the Brillouin zone and, eventually, enabling the possibilities for realizing non-trivial topological phases.
\twocolumngrid
\begin{figure*}[hbt!]
    \centering
    \begin{subfigure}{0.4\linewidth}
        \centering
         \caption[]{}
        \includegraphics[width=\linewidth]{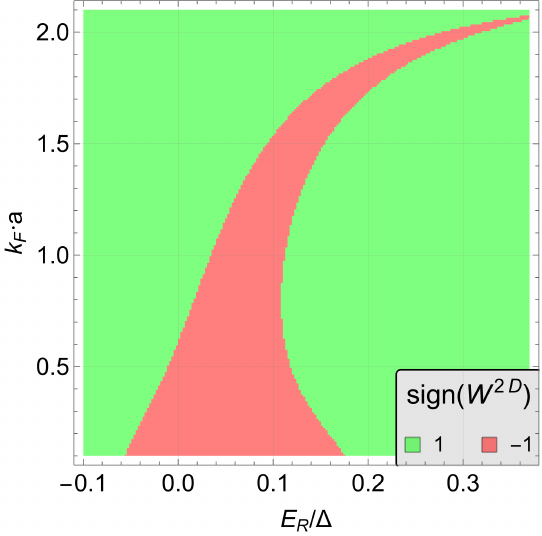}
        \label{Fig:2DPhaseDiagrammA}
    \end{subfigure}\hspace*{0.1\textwidth}
    \begin{subfigure}{0.41\linewidth}
        \centering
         \caption[]{}
        \includegraphics[width=\linewidth]
        {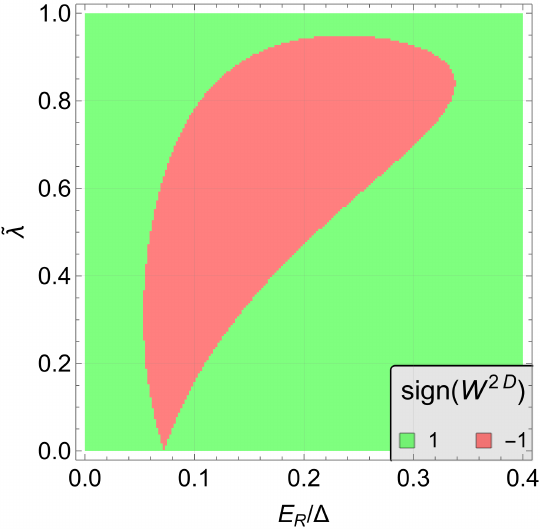}
        \label{Fig:2DPhaseDiagrammB}
    \end{subfigure}
    \caption{Topological phase diagram for two-dimensional lattices with periodic boundary conditions with a nearest-neighbor cutoff and a small repulsive on-site interaction $\tilde{U}$. Topological invariant, ${\rm sign} (W^{2D})$, depending (a) on the on-site energy $E_{\rm R}/\Delta$ and the Fermi wavevector $k_{\rm F}a$ for a fixed renormalized spin-orbit coupling strength $\tilde{\lambda}=0.25$ and (b) on the on-site energy $E_{\rm R}/\Delta$ and $\tilde{\lambda}$ for $k_{\rm F}a=1.2$. (parameters: $m=10^{-2}  (a^2 \Delta)^{-1}$, $\abs{V}=5\Delta$, and $\tilde{U}=0.05\Delta$.) }
    \label{Fig:2DPhaseDiagramm}
\end{figure*}
\section{TOPOLOGICAL INVARIANT IN 2D LATTICES}
\label{Appendix:TopologicalInvariant2D}
Here, we demonstrate the topological phases in 2D lattices by numerical computation of the topological invariant $W^{2D}$. A possible zero-energy crossing in one of the time-reversal inversion symmetric lines of the Brillouin zone $p_y=0$ or $p_y=\pi$ induces changes in the topological invariant $W^{2D}$. The topological phase diagram, see \cref{Fig:2DPhaseDiagramm}, can be directly compared to the 1D case of \cref{Fig.:1DTop} in the main text, that has been plotted for the same parameter set. Such a comparison indicates that the 2D geometry can support a non-trivial phase for a broader range of parameters. 
\section{FOURIER TRANSFORMATION OF MATRIX ELEMENTS IN DENSE 1D CHAINS} \label{Appendix:DenseChain} Here we focus on some qualitative characteristics of the plots in \cref{Fig.:Chains-Effective Parameters} and explain them by analyzing the continuous Fourier transform of the singlet, $\Delta_{i,j}^{S}$, and triplet, $\Delta_{i,j}^{T}$, superconducting matrix elements in \cref{eq:MatrixElements2D}. The continuous and discrete Fourier transforms are expected to match in the dense limit $k_{F}a\ll1$. The argument of the special functions of \cref{eq:MatrixElements2D} is approximately $\left( k_{F_{\pm}}+i/\xi \right)a\approx k_{F_{\pm}}a$, in the realistic limit $\xi\gg k_{F_{\pm}}$. For the above, we write the Fourier transforms of the relevant Bessel functions 
\begin{equation}\label{eq:ContFourierTransform}
\begin{split}
&\mathcal{F}_{0\pm}:=\mathcal{F}\left[ J_0(k_{F_{\pm}} \abs{j}) \right]=\dfrac{\Lambda_{\pm}(p)}{\sqrt{k^2_{F_{\pm}}-p^2}},\\ \
&\mathcal{F}_{1\pm}:=\mathcal{F}\left[\sgn(j) J_1(k_{F_{\pm}} \abs{j}) \right]=\dfrac{i p \Lambda_{\pm}(p)}{k_{F_{\pm}}\sqrt{k^2_{F_{\pm}}-p^2}},\\ &\Lambda_{\pm}(p)=\Theta(p+k_{F_{\pm}})-\Theta(p-k_{F_{\pm}}),
\end{split}
\end{equation}
where $\Theta$ is the Heaviside step function. Despite the considered approximations, the analytical expressions for the superconducting singlet $\Delta^{S}(p)\sim \Gamma_+ F_{0_{+}}+\Gamma_-F_{0_{-}}$ and triplet $\Delta^{T}(p)\sim \Gamma_+ F_{1_{+}}-\Gamma_-F_{1_{-}}$ pairings reveal the essential characteristics of \cref{Fig.:Chains-Effective Parameters} in the main text. In specific, \cref{eq:ContFourierTransform} predicts divergences at $p=k_{F_{\pm}}$ and a vanishing of the pairings for $\abs{p}> {\rm max}  \left( k_{F_{+}},k_{F_{-}}\right)$ due to the $\Lambda_{\pm}$ factor. These characteristics transcend to the discrete Fourier transform of the singlet and triplet order parameters that we analyze in the main text. Indeed, \cref{Fig.:Chains-Effective Parameters} reveals finite cusps at momenta $p=k_{F_{\pm}}$ and approximately flat regions, due to the finite cutoff considered, for momenta $\abs{p}> {\rm max}  \left( k_{F_{+}},k_{F_{-}}\right)$.

\bibliography{refs}

@article{PhysRevB.91.064505,
  title = {Topological {Yu}-{Shiba}-{Rusinov} chain from spin-orbit coupling},
  author = {Brydon, P. M. R. and Das Sarma, S. and Hui, Hoi-Yin and Sau, Jay D.},
  journal = {Phys. Rev. B},
  volume = {91},
  issue = {6},
  pages = {064505},
  numpages = {7},
  year = {2015},
  month = {Feb},
  publisher = {American Physical Society},
  doi = {10.1103/PhysRevB.91.064505}
}

@Article{10.21468/SciPostPhys.9.6.085,
	title={{Coupling quantum corrals to form artificial molecules}},
	author={Saoirsé E. Freeney and Samuel T. P. Borman andJacob W. Harteveld and Ingmar Swart},
	journal={SciPost Phys.},
	volume={9},
	pages={085},
	year={2020},
	publisher={SciPost},
	doi={10.21468/SciPostPhys.9.6.085}
}

@article{PhysRevB.110.134517,
  title = {Superconductivity in two-dimensional systems with unconventional Rashba bands},
  author = {Wang, Ran and Li, Jiayang and Huang, Xinliang and Wang, Lichuan and Song, Rui and Hao, Ning},
  journal = {Phys. Rev. B},
  volume = {110},
  issue = {13},
  pages = {134517},
  numpages = {14},
  year = {2024},
  month = {Oct},
  publisher = {American Physical Society},
  doi = {10.1103/PhysRevB.110.134517}
}

@article{Gomes2012,
author={Gomes, Kenjiro K.
and Mar, Warren
and Ko, Wonhee
and Guinea, Francisco
and Manoharan, Hari C.},
title={Designer Dirac fermions and topological phases in molecular graphene},
journal={Nature},
year={2012},
month={Mar},
day={01},
volume={483},
number={7389},
pages={306-310},
issn={1476-4687},
doi={10.1038/nature10941}
}

@article{
doi:10.1126/sciadv.aar5251,
author = {Howon Kim  and Alexandra Palacio-Morales  and Thore Posske  and Levente Rózsa  and Krisztián Palotás  and László Szunyogh  and Michael Thorwart  and Roland Wiesendanger },
title = {Toward tailoring {Majorana} bound states in artificially constructed magnetic atom chains on elemental superconductors},
journal = {Science Advances},
volume = {4},
number = {5},
pages = {eaar5251},
year = {2018},
doi = {10.1126/sciadv.aar5251}}

@article{doi:10.1021/acsnano.2c00467,
author = {Jolie, Wouter and Hung, Tzu-Chao and Niggli, Lorena and Verlhac, Benjamin and Hauptmann, Nadine and Wegner, Daniel and Khajetoorians, Alexander Ako},
title = {Creating Tunable Quantum Corrals on a {Rashba} Surface Alloy},
journal = {ACS Nano},
volume = {16},
number = {3},
pages = {4876-4883},
year = {2022},
doi = {10.1021/acsnano.2c00467}
}

@article{RevModPhys.78.373,
  title = {Impurity-induced states in conventional and unconventional superconductors},
  author = {Balatsky, A. V. and Vekhter, I. and Zhu, Jian-Xin},
  journal = {Rev. Mod. Phys.},
  volume = {78},
  issue = {2},
  pages = {373--433},
  numpages = {0},
  year = {2006},
  month = {May},
  publisher = {American Physical Society},
  doi = {10.1103/RevModPhys.78.373}
}

@article{Li2016 ,
    author={Li, Jian
    and Neupert, Titus
    and Wang, Zhijun
    and MacDonald, A. H.
    and Yazdani, A.
    and Bernevig, B. Andrei},
    title={Two-dimensional chiral topological superconductivity in {Shiba} lattices},
    journal={Nat. Commun},
    year={2016},
    month={Jul},
    day={28},
    volume={7},
    number={1},
    pages={12297},
    issn={2041-1723},
    doi={10.1038/ncomms12297}
}

@article{PhysRevB.88.155420,
  title = {Topological superconducting phase in helical {Shiba} chains},
  author = {Pientka, Falko and Glazman, Leonid I. and von Oppen, Felix},
  journal = {Phys. Rev. B},
  volume = {88},
  issue = {15},
  pages = {155420},
  numpages = {13},
  year = {2013},
  month = {Oct},
  publisher = {American Physical Society},
  doi = {10.1103/PhysRevB.88.155420}
}

@article{Schneider2023,
author={Schneider, Lucas
and Ton, Khai That
and Ioannidis, Ioannis
and Neuhaus-Steinmetz, Jannis
and Posske, Thore
and Wiesendanger, Roland
and Wiebe, Jens},
title={Proximity superconductivity in atom-by-atom crafted quantum dots},
journal={Nature},
year={2023},
month={Sep},
day={01},
volume={621},
number={7977},
pages={60-65},
issn={1476-4687},
doi={10.1038/s41586-023-06312-0}
}

@article{PhysRevB.79.245428,
  title = {Unconventional spin topology in surface alloys with Rashba-type spin splitting},
  author = {Mirhosseini, H. and Henk, J. and Ernst, A. and Ostanin, S. and Chiang, C.-T. and Yu, P. and Winkelmann, A. and Kirschner, J.},
  journal = {Phys. Rev. B},
  volume = {79},
  issue = {24},
  pages = {245428},
  numpages = {5},
  year = {2009},
  month = {Jun},
  publisher = {American Physical Society},
  doi = {10.1103/PhysRevB.79.245428}
}

@article{PhysRevB.76.073310,
  title = {Spin-orbit splitting in an anisotropic two-dimensional electron gas},
  author = {Premper, J. and Trautmann, M. and Henk, J. and Bruno, P.},
  journal = {Phys. Rev. B},
  volume = {76},
  issue = {7},
  pages = {073310},
  numpages = {4},
  year = {2007},
  month = {Aug},
  publisher = {American Physical Society},
  doi = {10.1103/PhysRevB.76.073310}
}

@article{PhysRevB.75.195414,
  title = {Enhanced Rashba spin-orbit splitting in {Bi} / {Ag}(111) and {Pb}/ {Ag}(111) surface alloys from first principles},
  author = {Bihlmayer, G. and Bl\"ugel, S. and Chulkov, E. V.},
  journal = {Phys. Rev. B},
  volume = {75},
  issue = {19},
  pages = {195414},
  numpages = {6},
  year = {2007},
  month = {May},
  publisher = {American Physical Society},
  doi = {10.1103/PhysRevB.75.195414}
}

@article{PhysRevLett.98.186807,
  title = {Giant Spin Splitting through Surface Alloying},
  author = {Ast, Christian R. and Henk, J\"urgen and Ernst, Arthur and Moreschini, Luca and Falub, Mihaela C. and Pacil\'e, Daniela and Bruno, Patrick and Kern, Klaus and Grioni, Marco},
  journal = {Phys. Rev. Lett.},
  volume = {98},
  issue = {18},
  pages = {186807},
  numpages = {4},
  year = {2007},
  month = {May},
  publisher = {American Physical Society},
  doi = {10.1103/PhysRevLett.98.186807}
}

@book{Ashcroft76,
  author = {Ashcroft, N. W. and Mermin, N. D.},
  publisher = {Holt-Saunders},
  title = {{S}olid {S}tate {P}hysics},
  year = 1976
}

@article{PhysRevB.94.161110,
  title = {No-go theorem for a time-reversal invariant topological phase in noninteracting systems coupled to conventional superconductors},
  author = {Haim, Arbel and Berg, Erez and Flensberg, Karsten and Oreg, Yuval},
  journal = {Phys. Rev. B},
  volume = {94},
  issue = {16},
  pages = {161110},
  numpages = {5},
  year = {2016},
  month = {Oct},
  publisher = {American Physical Society},
  doi = {10.1103/PhysRevB.94.161110}
}

@article{PhysRevB.111.144508,
  title = {Classification of pair symmetries in superconductors with unconventional magnetism},
  author = {Maeda, Kazuki and Fukaya, Yuri and Yada, Keiji and Lu, Bo and Tanaka, Yukio and Cayao, Jorge},
  journal = {Phys. Rev. B},
  volume = {111},
  issue = {14},
  pages = {144508},
  numpages = {14},
  year = {2025},
  month = {Apr},
  publisher = {American Physical Society},
  doi = {10.1103/PhysRevB.111.144508},
}

@Article{Su2017,
author={Su, Zhaoen
and Tacla, Alexandre B.
and Hocevar, Mo{\"i}ra
and Car, Diana
and Plissard, S{\'e}bastien R.
and Bakkers, Erik P. A. M.
and Daley, Andrew J.
and Pekker, David
and Frolov, Sergey M.},
title={Andreev molecules in semiconductor nanowire double quantum dots},
journal={Nat. Comm.},
year={2017},
month={Sep},
day={19},
volume={8},
number={1},
pages={585},
issn={2041-1723},
doi={10.1038/s41467-017-00665-7},
url={https://doi.org/10.1038/s41467-017-00665-7}
}

@Article{Janvier2015,
author={Janvier, C.
and Tosi, L.
and Bretheau, L.
and Girit, {\c{C}} {\"O}
and Stern, M.
and Bertet, P.
and Joyez, P.
and Vion, D.
and Esteve, D.
and Goffman, M. F.
and Pothier, H.
and Urbina, C.},
title={Coherent manipulation of {Andreev} states in superconducting atomic contacts},
journal={Science},
year={2015},
month={Sep},
day={11},
publisher={American Association for the Advancement of Science},
volume={349},
number={6253},
pages={1199-1202},
doi={10.1126/science.aab2179},
url={https://doi.org/10.1126/science.aab2179}
}

@article{
doi:10.1126/science.adf2685,
author = {E. Sierda  and X. Huang  and D. I. Badrtdinov  and B. Kiraly  and E. J. Knol  and G. C. Groenenboom  and M. I. Katsnelson  and M. Rösner  and D. Wegner  and A. A. Khajetoorians },
title = {Quantum simulator to emulate lower-dimensional molecular structure},
journal = {Science},
volume = {380},
number = {6649},
pages = {1048-1052},
year = {2023},
doi = {10.1126/science.adf2685},
eprint = {https://www.science.org/doi/pdf/10.1126/science.adf2685}}

@article{RevModPhys.88.035005,
  title = {Classification of topological quantum matter with symmetries},
  author = {Chiu, Ching-Kai and Teo, Jeffrey C. Y. and Schnyder, Andreas P. and Ryu, Shinsei},
  journal = {Rev. Mod. Phys.},
  volume = {88},
  issue = {3},
  pages = {035005},
  numpages = {63},
  year = {2016},
  month = {Aug},
  publisher = {American Physical Society},
  doi = {10.1103/RevModPhys.88.035005},
  url = {https://link.aps.org/doi/10.1103/RevModPhys.88.035005}
}

@article{PhysRevB.88.134523,
  title = {Topological invariant for generic one-dimensional time-reversal-symmetric superconductors in class {DIII}},
  author = {Budich, Jan Carl and Ardonne, Eddy},
  journal = {Phys. Rev. B},
  volume = {88},
  issue = {13},
  pages = {134523},
  numpages = {5},
  year = {2013},
  month = {Oct},
  publisher = {American Physical Society},
  doi = {10.1103/PhysRevB.88.134523},
  url = {https://link.aps.org/doi/10.1103/PhysRevB.88.134523}
}

@article{PhysRevLett.77.3419,
  title = {Spin Splitting of an {Au}(111) Surface State Band Observed with Angle Resolved Photoelectron Spectroscopy},
  author = {LaShell, S. and McDougall, B. A. and Jensen, E.},
  journal = {Phys. Rev. Lett.},
  volume = {77},
  issue = {16},
  pages = {3419--3422},
  numpages = {0},
  year = {1996},
  month = {Oct},
  publisher = {American Physical Society},
  doi = {10.1103/PhysRevLett.77.3419}
}

@article{PhysRevLett.114.106801,
  title = {Strong Localization of {Majorana} End States in Chains of Magnetic Adatoms},
  author = {Peng, Yang and Pientka, Falko and Glazman, Leonid I. and von Oppen, Felix},
  journal = {Phys. Rev. Lett.},
  volume = {114},
  issue = {10},
  pages = {106801},
  numpages = {5},
  year = {2015},
  month = {Mar},
  publisher = {American Physical Society},
  doi = {10.1103/PhysRevLett.114.106801},
  url = {https://link.aps.org/doi/10.1103/PhysRevLett.114.106801}
}

@article{Wang_2013,
doi = {10.1088/0953-8984/25/15/155601},
url = {https://dx.doi.org/10.1088/0953-8984/25/15/155601},
year = {2013},
month = {mar},
publisher = {IOP Publishing},
volume = {25},
number = {15},
pages = {155601},
author = {Zhong Wang and Binghai Yan},
title = {Topological {Hamiltonian} as an exact tool for topological invariants},
journal = {J. Phys.:Condens. Matter}
}

@ARTICLE{1972PThPh..47.1817M,
       author = {{Machida}, K. and {Shibata}, F.},
        title = {Bound States Due to Resonance Scattering in Superconductor},
      journal = {Prog. Theor. Phys.},
         year = 1972,
        month = jun,
       volume = {47},
       number = {6},
        pages = {1817-1823},
          doi = {10.1143/PTP.47.1817},
       adsurl = {https://ui.adsabs.harvard.edu/abs/1972PThPh..47.1817M}
}

@article{PhysRevB.110.L100505,
  title = {Scanning tunneling spectroscopy study of proximity superconductivity in finite-size quantized surface states},
  author = {Schneider, Lucas and von Bredow, Christian and Kim, Howon and That Ton, Khai and H\"anke, Torben and Wiebe, Jens and Wiesendanger, Roland},
  journal = {Phys. Rev. B},
  volume = {110},
  issue = {10},
  pages = {L100505},
  numpages = {6},
  year = {2024},
  month = {Sep},
  publisher = {American Physical Society},
  doi = {10.1103/PhysRevB.110.L100505},
  url = {https://link.aps.org/doi/10.1103/PhysRevB.110.L100505}
}

@article{Yu1965Bound,
 title={Bound state in superconductors with paramagnetic impurities},
  author={{Yu}, L},
  journal={Acta Phys. Sin.},
  volume={21},
  number = {1},
  pages={75--91},
  year={1965},
  doi ={doi: 10.7498/aps.21.75},
  publisher={Chinese Physical Society}
}

@article{Shiba1968classical,
  title={Classical spins in superconductors},
  author={Shiba, Hiro{Yu}ki},
  journal={Progr. Theoret. Phys.},
  volume={40},
  number={3},
  pages={435--451},
  year={1968},
  url = {https://doi.org/10.1143/PTP.40.435},
  doi = {doi.org/10.1143/PTP.40.435},
  publisher={Oxford University Press}
}

@article{HATTAB20088227,
title = {Epitaxial {Bi}(111) films on {Si}(001): Strain state, surface morphology, and defect structure},
journal = {Thin Solid Films},
volume = {516},
number = {23},
pages = {8227-8231},
year = {2008},
issn = {0040-6090},
doi = {https://doi.org/10.1016/j.tsf.2008.02.038},
author = {H. Hattab and E. Zubkov and A. Bernhart and G. Jnawali and C. Bobisch and B. Krenzer and M. Acet and R. Möller and M. {Horn-von Hoegen}}
}

@article{Rusinov1969theory,
  title={On the theory of gapless superconductivity in alloys containing paramagnetic impurities},
  author={{Rusinov}, A I},
  journal={J. Exp. Theor. Phys.},
  volume={29},
  issue ={6},
  pages={1101},
  year={1969}
}

@article{PhysRevB.104.245133,
  title = {Subgap states at ferromagnetic and spiral-ordered magnetic chains in two-dimensional superconductors. {I}. Continuum description},
  author = {Carroll, C. J. F. and Braunecker, B.},
  journal = {Phys. Rev. B},
  volume = {104},
  issue = {24},
  pages = {245133},
  numpages = {13},
  year = {2021},
  month = {Dec},
  publisher = {American Physical Society},
  doi = {10.1103/PhysRevB.104.245133},
  url = {https://link.aps.org/doi/10.1103/PhysRevB.104.245133}
}

@Article{Ton2025,
author={Ton, Khai That
and Xu, Chang
and Ioannidis, Ioannis
and Schneider, Lucas
and Posske, Thore
and Wiesendanger, Roland
and Morr, Dirk K.
and Wiebe, Jens},
title={Non-local detection of coherent Yu--Shiba--Rusinov quantum projections},
journal={Nature Physics},
year={2025},
month={Nov},
day={26},
issn={1745-2481},
doi={10.1038/s41567-025-03109-y},
url={https://doi.org/10.1038/s41567-025-03109-y}
}

@article{PhysRevB.78.195125,
  title = {Classification of topological insulators and superconductors in three spatial dimensions},
  author = {Schnyder, Andreas P. and Ryu, Shinsei and Furusaki, Akira and Ludwig, Andreas W. W.},
  journal = {Phys. Rev. B},
  volume = {78},
  issue = {19},
  pages = {195125},
  numpages = {22},
  year = {2008},
  month = {Nov},
  publisher = {American Physical Society},
  doi = {10.1103/PhysRevB.78.195125},
  url = {https://link.aps.org/doi/10.1103/PhysRevB.78.195125}
}

@article{10.1063/1.3149495,
    author = {Kitaev, Alexei},
    title = {Periodic table for topological insulators and superconductors},
    journal = {AIP Conf. Proc.},
    volume = {1134},
    number = {1},
    pages = {22-30},
    year = {2009},
    month = {05},
    issn = {0094-243X},
    doi = {10.1063/1.3149495},
    url = {https://doi.org/10.1063/1.3149495},
}

@book{8857dd78fb874b8ca4eb9995a2671dfc,
title = "Many-body quantum theory in condensed matter physics - an introduction",
author = "Henrik Bruus and Karsten Flensberg",
year = "2004",
publisher = "Oxford University Press",
address = "United States",
}

@book{Coleman_2015, place={Cambridge}, title={Introduction to Many-Body Physics}, publisher={Cambridge University Press}, author={Coleman, Piers}, year={2015}}

@article{PhysRevLett.124.246803,
  title = {Suppressed Charge Dispersion via Resonant Tunneling in a Single-Channel Transmon},
  author = {Kringh\o{}j, A. and van Heck, B. and Larsen, T. W. and Erlandsson, O. and Sabonis, D. and Krogstrup, P. and Casparis, L. and Petersson, K. D. and Marcus, C. M.},
  journal = {Phys. Rev. Lett.},
  volume = {124},
  issue = {24},
  pages = {246803},
  numpages = {6},
  year = {2020},
  month = {Jun},
  publisher = {American Physical Society},
  doi = {10.1103/PhysRevLett.124.246803},
  url = {https://link.aps.org/doi/10.1103/PhysRevLett.124.246803}
}

@article{PhysRevLett.124.246802,
  title = {Observation of Vanishing Charge Dispersion of a Nearly Open Superconducting Island},
  author = {Bargerbos, Arno and Uilhoorn, Willemijn and Yang, Chung-Kai and Krogstrup, Peter and Kouwenhoven, Leo P. and de Lange, Gijs and van Heck, Bernard and Kou, Angela},
  journal = {Phys. Rev. Lett.},
  volume = {124},
  issue = {24},
  pages = {246802},
  numpages = {7},
  year = {2020},
  month = {Jun},
  publisher = {American Physical Society},
  doi = {10.1103/PhysRevLett.124.246802},
  url = {https://link.aps.org/doi/10.1103/PhysRevLett.124.246802}
}

@article{PhysRevB.96.161407,
  title = {{DIII} topological superconductivity with emergent time-reversal symmetry},
  author = {Reeg, Christopher and Schrade, Constantin and Klinovaja, Jelena and Loss, Daniel},
  journal = {Phys. Rev. B},
  volume = {96},
  issue = {16},
  pages = {161407},
  numpages = {5},
  year = {2017},
  month = {Oct},
  publisher = {American Physical Society},
  doi = {10.1103/PhysRevB.96.161407},
  url = {https://link.aps.org/doi/10.1103/PhysRevB.96.161407}
}

@article{Zubarev,
doi = {10.1070/PU1960v003n03ABEH003275},
url = {https://dx.doi.org/10.1070/PU1960v003n03ABEH003275},
year = {1960},
month = {mar},
publisher = {},
volume = {3},
number = {3},
pages = {320},
author = {D N Zubarev},
title = {DOUBLE-TIME {GREEN} FUNCTIONS IN STATISTICAL PHYSICS},
journal = {Soviet Physics Uspekhi}
}

@article{PhysRev.56.317,
  title = {On the Surface States Associated with a Periodic Potential},
  author = {Shockley, William},
  journal = {Phys. Rev.},
  volume = {56},
  issue = {4},
  pages = {317--323},
  numpages = {0},
  year = {1939},
  month = {Aug},
  publisher = {American Physical Society},
  doi = {10.1103/PhysRev.56.317},
  url = {https://link.aps.org/doi/10.1103/PhysRev.56.317}
}

@article{Bazarnik2023,
author={Bazarnik, Maciej
and Lo Conte, Roberto
and Mascot, Eric
and von Bergmann, Kirsten
and Morr, Dirk K.
and Wiesendanger, Roland},
journal={Nat. Commun.},
title={Antiferromagnetism-driven two-dimensional topological nodal-point superconductivity},
year={2023},
month={Feb},
day={04},
volume={14},
number={1},
pages={614},
issn={2041-1723},
doi={10.1038/s41467-023-36201-z},
url={https://doi.org/10.1038/s41467-023-36201-z}
}

@article{
doi:10.1126/science.1259327,
author = {Stevan Nadj-Perge  and Ilya K. Drozdov  and Jian Li  and Hua Chen  and Sangjun Jeon  and Jungpil Seo  and Allan H. MacDonald  and B. Andrei Bernevig  and Ali Yazdani },
title = {Observation of {Majorana} fermions in ferromagnetic atomic chains on a superconductor},
journal = {Science},
volume = {346},
number = {6209},
pages = {602-607},
year = {2014},
doi = {10.1126/science.1259327},
url = {https://www.science.org/doi/abs/10.1126/science.1259327}}

@Article{Ménard2017,
author={M{\'e}nard, Gerbold C.
and Guissart, S{\'e}bastien
and Brun, Christophe
and Leriche, Rapha{\"e}l T.
and Trif, Mircea
and Debontridder, Fran{\c{c}}ois
and Demaille, Dominique
and Roditchev, Dimitri
and Simon, Pascal
and Cren, Tristan},
title={Two-dimensional topological superconductivity in {Pb}/{Co}/{Si}(111)},
journal={Nat. Commun.},
year={2017},
month={Dec},
day={11},
volume={8},
number={1},
pages={2040},
issn={2041-1723},
doi={10.1038/s41467-017-02192-x},
url={https://doi.org/10.1038/s41467-017-02192-x}
}

@article{
doi:10.1126/sciadv.aav6600,
author = {Alexandra Palacio-Morales  and Eric Mascot  and Sagen Cocklin  and Howon Kim  and Stephan Rachel  and Dirk K. Morr  and Roland Wiesendanger },
title = {Atomic-scale interface engineering of {Majorana} edge modes in a 2D magnet-superconductor hybrid system},
journal = {Sci. Adv.},
volume = {5},
number = {7},
pages = {eaav6600},
year = {2019},
doi = {10.1126/sciadv.aav6600},
URL = {https://www.science.org/doi/abs/10.1126/sciadv.aav6600}}

@article{Heimes_2015,
doi = {10.1088/1367-2630/17/2/023051},
url = {https://dx.doi.org/10.1088/1367-2630/17/2/023051},
year = {2015},
month = {feb},
publisher = {IOP Publishing},
volume = {17},
number = {2},
pages = {023051},
author = {Heimes, Andreas and Mendler, Daniel and Kotetes, Panagiotis},
title = {Interplay of topological phases in magnetic adatom-chains on top of a {Rashba} superconducting surface},
journal = {New J. Phys.}
}

@article{PhysRevLett.121.196801,
  title = {{Majorana} {Kramers} Pairs in Higher-Order Topological Insulators},
  author = {Hsu, Chen-Hsuan and Stano, Peter and Klinovaja, Jelena and Loss, Daniel},
  journal = {Phys. Rev. Lett.},
  volume = {121},
  issue = {19},
  pages = {196801},
  numpages = {8},
  year = {2018},
  month = {Nov},
  publisher = {American Physical Society},
  doi = {10.1103/PhysRevLett.121.196801},
  url = {https://link.aps.org/doi/10.1103/PhysRevLett.121.196801}
}

@article{Kitaev2001,
title = {Unpaired {Majorana} fermions in quantum
wires},
doi = {10.1070/1063-7869/44/10S/S29},
url = {https://dx.doi.org/10.1070/1063-7869/44/10S/S29},
year = {2001},
month = {oct},
publisher = {},
volume = {44},
number = {10S},
pages = {131},
author = {A Yu Kitaev},
journal = {Phys.--Usp.}
}

@article{Leijnse_2012,
doi = {10.1088/0268-1242/27/12/124003},
url = {https://dx.doi.org/10.1088/0268-1242/27/12/124003},
year = {2012},
month = {nov},
publisher = {IOP Publishing},
volume = {27},
number = {12},
pages = {124003},
author = {Leijnse, Martin and Flensberg, Karsten},
title = {Introduction to topological superconductivity and {Majorana} fermions},
journal = {Semicond. Sci. Technol.}
}

@article{Sarma2015,
author={Sarma, Sankar Das
and Freedman, Michael
and Nayak, Chetan},
title={{Majorana} zero modes and topological quantum computation},
journal={Npj Quantum Inf.},
year={2015},
month={Oct},
day={27},
volume={1},
number={1},
pages={15001},
issn={2056-6387},
doi={10.1038/npjqi.2015.1},
url={https://doi.org/10.1038/npjqi.2015.1}
}

@article{PhysRevLett.105.077001,
  title = {{Majorana} Fermions and a Topological Phase Transition in Semiconductor-Superconductor Heterostructures},
  author = {Lutchyn, Roman M. and Sau, Jay D. and Das Sarma, S.},
  journal = {Phys. Rev. Lett.},
  volume = {105},
  issue = {7},
  pages = {077001},
  numpages = {4},
  year = {2010},
  month = {Aug},
  publisher = {American Physical Society},
  doi = {10.1103/PhysRevLett.105.077001},
  url = {https://link.aps.org/doi/10.1103/PhysRevLett.105.077001}
}

@article{PhysRevLett.105.177002,
  title = {Helical Liquids and {Majorana} Bound States in Quantum Wires},
  author = {Oreg, Yuval and Refael, Gil and von Oppen, Felix},
  journal = {Phys. Rev. Lett.},
  volume = {105},
  issue = {17},
  pages = {177002},
  numpages = {4},
  year = {2010},
  month = {Oct},
  publisher = {American Physical Society},
  doi = {10.1103/PhysRevLett.105.177002},
  url = {https://link.aps.org/doi/10.1103/PhysRevLett.105.177002}
}

@article{PhysRevLett.100.096407,
  title = {Superconducting Proximity Effect and Majorana Fermions at the Surface of a Topological Insulator},
  author = {Fu, Liang and Kane, C. L.},
  journal = {Phys. Rev. Lett.},
  volume = {100},
  issue = {9},
  pages = {096407},
  numpages = {4},
  year = {2008},
  month = {Mar},
  publisher = {American Physical Society},
  doi = {10.1103/PhysRevLett.100.096407},
  url = {https://link.aps.org/doi/10.1103/PhysRevLett.100.096407}
}

@article{RevModPhys.80.1083,
  title = {Non-Abelian anyons and topological quantum computation},
  author = {Nayak, Chetan and Simon, Steven H. and Stern, Ady and Freedman, Michael and Das Sarma, Sankar},
  journal = {Rev. Mod. Phys.},
  volume = {80},
  issue = {3},
  pages = {1083--1159},
  numpages = {0},
  year = {2008},
  month = {Sep},
  publisher = {American Physical Society},
  doi = {10.1103/RevModPhys.80.1083},
  url = {https://link.aps.org/doi/10.1103/RevModPhys.80.1083}
}

@article{
doi:10.1126/science.1222360,
author = {V. Mourik  and K. Zuo  and S. M. Frolov  and S. R. Plissard  and E. P. A. M. Bakkers  and L. P. Kouwenhoven },
title = {Signatures of {Majorana} Fermions in Hybrid Superconductor-Semiconductor Nanowire Devices},
journal = {Science},
volume = {336},
number = {6084},
pages = {1003-1007},
year = {2012},
doi = {10.1126/science.1222360},
URL = {https://www.science.org/doi/abs/10.1126/science.1222360}}

@article{Das2012,
author={Das, Anindya
and Ronen, Yuval
and Most, Yonatan
and Oreg, Yuval
and Heiblum, Moty
and Shtrikman, Hadas},
title={Zero-bias peaks and splitting in an {A}l--{I}n{A}s nanowire topological superconductor as a signature of {Majorana} fermions},
journal={Nature Physics},
year={2012},
month={Dec},
day={01},
volume={8},
number={12},
pages={887-895},
issn={1745-2481},
doi={10.1038/nphys2479},
url={https://doi.org/10.1038/nphys2479}
}

@article{Zhang2021,
author={Zhang, Hao
and Liu, Chun-Xiao
and Gazibegovic, Sasa
and Xu, Di
and Logan, John A.
and Wang, Guanzhong
and van Loo, Nick
and Bommer, Jouri D. S.
and de Moor, Michiel W. A.
and Car, Diana
and Op het Veld, Roy L. M.
and van Veldhoven, Petrus J.
and Koelling, Sebastian
and Verheijen, Marcel A.
and Pendharkar, Mihir
and Pennachio, Daniel J.
and Shojaei, Borzoyeh
and Lee, Joon Sue
and Palmstr{\o}m, Chris J.
and Bakkers, Erik P. A. M.
and Das Sarma, S.
and Kouwenhoven, Leo P.},
title={Retraction Note: Quantized {Majorana} conductance},
journal={Nature},
year={2021},
month={Mar},
day={01},
volume={591},
number={7851},
pages={E30-E30},
issn={1476-4687},
doi={10.1038/s41586-021-03373-x},
url={https://doi.org/10.1038/s41586-021-03373-x}
}

@article{PhysRevB.105.174519,
  title = {Quasiparticle poisoning in trivial and topological {Josephson} junctions},
  author = {Svetogorov, Aleksandr E. and Loss, Daniel and Klinovaja, Jelena},
  journal = {Phys. Rev. B},
  volume = {105},
  issue = {17},
  pages = {174519},
  numpages = {11},
  year = {2022},
  month = {May},
  publisher = {American Physical Society},
  doi = {10.1103/PhysRevB.105.174519},
  url = {https://link.aps.org/doi/10.1103/PhysRevB.105.174519}
}

@article{Liebhaber2022,
author={Liebhaber, Eva
and R{\"u}tten, Lisa M.
and Reecht, Ga{\"e}l
and Steiner, Jacob F.
and Rohlf, Sebastian
and Rossnagel, Kai
and von Oppen, Felix
and Franke, Katharina J.},
title={Quantum spins and hybridization in artificially-constructed chains of magnetic adatoms on a superconductor},
journal={Nat. Commun.},
year={2022},
month={Apr},
day={20},
volume={13},
number={1},
pages={2160},
issn={2041-1723},
doi={10.1038/s41467-022-29879-0},
url={https://doi.org/10.1038/s41467-022-29879-0}
}

@article{PhysRevB.108.L220506,
  title = {Surrogate model solver for impurity-induced superconducting subgap states},
  author = {Baran, Virgil V. and Frost, Emil J. P. and Paaske, Jens},
  journal = {Phys. Rev. B},
  volume = {108},
  issue = {22},
  pages = {L220506},
  numpages = {6},
  year = {2023},
  month = {Dec},
  publisher = {American Physical Society},
  doi = {10.1103/PhysRevB.108.L220506},
  url = {https://link.aps.org/doi/10.1103/PhysRevB.108.L220506}
}

@Article{Bihlmayer2022,
author={Bihlmayer, Gustav
and No{\"e}l, Paul
and Vyalikh, Denis V.
and Chulkov, Evgueni V.
and Manchon, Aur{\'e}lien},
title={Rashba-like physics in condensed matter},
journal={Nature Reviews Physics},
year={2022},
month={Oct},
day={01},
volume={4},
number={10},
pages={642-659},
issn={2522-5820},
doi={10.1038/s42254-022-00490-y},
url={https://doi.org/10.1038/s42254-022-00490-y}
}

\end{document}